\documentclass{jfm}
\usepackage{graphicx}
\usepackage{epstopdf, epsfig}
\usepackage{subfigure}
\usepackage{epsf}
\usepackage{bm}
\usepackage{multirow}
\usepackage{float}
\usepackage{color}
\usepackage{mathrsfs}

\shorttitle{Deformable bubbles rising initially in line}
\shortauthor{J. Zhang, M. J. Ni, J. Magnaudet}

\title{Three-dimensional dynamics of a pair of deformable bubbles  rising initially in line. Part 1: Moderately inertial regimes}

\author{Jie Zhang\aff{1}, 
Ming-Jiu Ni\aff{1,2}
\corresp{\email{mjni@ucas.ac.cn}} 
\and Jacques Magnaudet\aff{3}
\corresp{\email{Jacques.Magnaudet@imft.fr}}}

\affiliation{\aff{1}State Key Laboratory for Strength and Vibration of Mechanical Structures, School of Aerospace, Xi'an Jiaotong University, Xi'an, China
\aff{2}School of Engineering, University of Chinese Academy of Sciences, Beijing, China
\aff{3}Institut de M\'ecanique des Fluides de Toulouse (IMFT), Universit\'e de Toulouse, CNRS, Toulouse, France}

\begin{document}

\maketitle

\begin{abstract}

The buoyancy-driven motion of two identical gas bubbles released in line in a liquid at rest is examined with the help of highly resolved simulations, focusing on moderately inertial regimes in which the path of an isolated bubble is vertical.  Assuming first an axisymmetric evolution, equilibrium configurations of the bubble pair are determined as a function of the buoyancy-to-viscous and buoyancy-to-capillary force ratios which define the Galilei ($Ga$) and Bond ($Bo$) numbers of the system, respectively.  
The three-dimensional solutions reveal that this axisymmetric equilibrium is actually never reached. Instead, provided $Bo$ stands below a critical $Ga$-dependent threshold determining the onset of coalescence, two markedly different evolutions are observed. At the lower end of the explored $(Ga,\,Bo)$-range, the tandem follows a Drafting-Kissing-Tumbling scenario which eventually yields a planar side-by-side motion. For larger $Ga$, the trailing bubble drifts laterally and gets out of the wake of the leading bubble, barely altering the path of the latter. In this second scenario, the late configuration is characterized by a significant inclination of the tandem ranging from $30^\circ$ to $40^\circ$ with respect to the vertical. Bubble deformation has a major influence on the evolution of the system. It controls the magnitude of vortical effects in the wake of the leading bubble, hence the strength of the attractive force acting on the trailing bubble. It also governs the strength and even the sign of the lateral force acting on this bubble, a mechanism of particular importance when the tandem is released with a small angular deviation.
\end{abstract}

\section{Introduction}\label{sec1}

Buoyancy-driven bubbly flows are widely encountered in natural environments (\textit{e.g.} breaking waves, bubbly plumes released from the floor of lakes and oceans) and engineering devices (\textit{e.g.} bubble columns, ladle steel making, boiling flows in power plants). Such applications have driven fundamental studies of fluid-bubble interactions in bubbly suspensions for a long time. 
Early computational investigations \citep{Smereka1993,Sangani1993} assumed spherical bubble shapes and neglected any possible influence of vorticity in the liquid. The corresponding potential flow simulations predict the formation of large horizontal bubble clusters. However, subsequent laboratory experiments and simulations based on the full Navier-Stokes equations revealed a less clear-cut picture. When bubbles remain nearly spherical and viscous effects, while smaller than inertial effects, remain significant in the bulk, experiments \citep{Cartellier2001} and three-dimensional simulations \citep{Esmaeeli1998,bunner2002dynamics, Yin2008,loisy2017buoyancy} show that the microstucture is governed by the pair interaction mechanism known as  Drafting-Kissing-Tumbling (hereinafter abbreviated as DKT) for sedimenting solid particles \citep{joseph1986nonlinear, fortes1987nonlinear}. This is primarily a wake effect by which two particles or bubbles initially aligned vertically are first attracted toward each other, then repel in the horizontal direction when they get very close to each other, until they reach an equilibrium separation and fall/rise side by side. The process is self-repeating, since at some point each of the two bodies enters unavoidably in the wake of one of its neighbours. In more inertial regimes, experiments \citep{Cartellier2001,Zenit2001,Figueroa2005} and simulations \citep{Esmaeeli1999,Yin2008} with nearly-spherical bubbles reveal a clear tendency of bubbles to align horizontally. However, the corresponding clusters are less strong, \textit{i.e.} the bubble distribution is less anisotropic, than predicted by potential flow theory. Simulations also considered effects of bubble deformation. In moderately inertial regimes, the results show that pairs of significantly oblate bubbles tend to align vertically, forming vertical streams \citep{bunner2003effect}. However this ` chimney' effect disappears in strongly inertial regimes in which bubbles tend to follow zigzagging or spiralling paths  \citep{Esmaeeli2005}.\vspace{1mm}\\
\indent This brief review highlights the fact that the microstructure of buoyancy-driven bubbly suspensions is to a large extent governed by pair interactions. In particular, the two canonical configurations in which two bubbles are released either in line or side by side are of particular relevance to obtain a better understanding of the local mechanisms at stake in freely-rising suspensions. Configurations corresponding to intermediate initial inclinations connect these two extreme geometries and were considered in the nearly-inviscid limit, both theoretically (assuming a spherical bubble shape) and experimentally, by \cite{Kok1993,Kok1993b}. 
Detailed experiments were carried out with bubble pairs rising side by side \citep{Duineveld1998,Sanada2009,kong2019hydrodynamic}, varying the liquid properties, bubble sizes and initial separation. Depending on flow conditions, the two bubbles were found to repel or attract each other. In the latter case, they may reach an equilibrium separation or collide, in which case they subsequently bounce or coalesce. 
The respective roles of irrotational and vortical effects on the sign and magnitude of the transverse interaction force were examined in the simulations of \cite{Legendre2003} assuming spherical bubbles. In particular, a regime map predicting the characteristics of the final configuration as a function of the initial separation was obtained. 
Influence of bubble deformation, which beyond a critical oblateness makes the wake unstable, was considered numerically by \cite{zhang2019vortex}. The resulting double-threaded wakes and their interactions were found to be critically important during the collision stages. Indeed, in most cases this interaction is responsible for an extra repulsive transverse force which makes the two bubbles bounce.\\
\indent Not surprisingly, the in-line configuration was first considered  assuming spherical bubbles and an axisymmetric flow at all time. With an irrotational flow in the bulk supplemented with a weak boundary layer and wake past each bubble, \cite{Harper1970} established the existence of a finite equilibrium separation of the two bubbles. He showed that this equilibrium stems from the balance between a repulsive force corresponding to the irrotational flow past the two bodies and an attractive force resulting from the influence of the boundary layer past the leading bubble on the boundary layer of the trailing bubble. His conclusions were qualitatively confirmed and extended toward lower Reynolds numbers through axisymmetric computations by \cite{Yuan1994} (and later by \cite{hallez2011interaction} who considered arbitrary orientations of the bubble pair). This investigation prompted Harper to improve his theory by accounting for viscous diffusion in the wake of the leading bubble, allowing him to reach a better agreement with the numerical results for large Reynolds numbers \citep{Harper1997}. Early experiments with nearly spherical bubbles were performed in weaker inertial regimes corresponding to Reynolds numbers lower than those considered by \cite{Yuan1994}. Using distilled water, \cite{katz1996wake} observed that under such conditions the two bubbles always collide and coalesce. Experiments performed in silicone oils by \cite{watanabe2006line} in the same regime confirmed that the bubbles collide but revealed no coalescence. 
In moderately inertial regimes, these authors did not observe any head-on collision, in line with the numerical findings of \cite{Yuan1994}. However, they found the equilibrium axisymmetric configuration to be unstable, confirming Harper's theoretical analysis \citep{Harper1970}. The three-dimensional evolution of the bubble pair in moderately inertial regimes was explored in more detail by \cite{Kusuno2015} and \cite{Kusuno2019}, using ultrapure water and silicone oil, respectively. Several interaction scenarios were reported, including the DKT process and a distinct evolution (also noticed in the computations of \cite{gumulya2017interaction}) in which the trailing bubble drifts laterally without significantly modifying the path of the leading bubble. In this case, the separation between the two bubbles remained significantly larger than their radius throughout the rise.  \vspace{1mm}\\
\indent Focusing on the initial in-line configuration, possibly with some small angular deviation, the present investigation aims at providing a more detailed understanding of the interaction processes reviewed above. For this purpose, we carried out high-resolution three-dimensional time-dependent computations allowing a complete interplay of inertial, viscous and capillary effects over a wide range of flow regimes.
The present paper reports on the first half of this investigation. It focuses on moderately inertial regimes in which each bubble, taken isolated, would follow a straight vertical path. High-inertia regimes in which isolated bubbles follow a non-straight path will be examined in the companion paper (\cite{jiesubmitted}). 
The entire work is based on the open source code \textit{Basilisk} \citep{popinet2015quadtree} which employs the volume of fluid (VOF) approach. This method enables the bubbles to deform freely in the liquid as they rise and interact. Moreover, the adaptive mesh refinement (AMR) technique embedded in this code, supplemented with a specific refinement \citep{zhang2019vortex}, makes it possible to properly capture the flow in the vicinity of the bubble surface, in the near wake, as well as within the film that forms between the two bubbles when they get very close to each other. 
The paper is organized as follows. The problem and the dimensionless parameters are introduced in \S\,2, while \S\,3 summarizes the numerical method. Before embarking on the discussion of numerical results, the fundamental mechanisms involved in the interaction process for spherical and deformed bubbles are reviewed in \S\,\ref{mechanisms}. Predictions obtained by constraining the flow to remain axisymmetric are discussed in \S\,\ref{sec4.1}. Then, fully three-dimensional evolutions are examined in \S\,\ref{3D}. Influence of bubble deformation and initial conditions on the evolution of the bubble pair is discussed in \S\,\ref{sec5}. The main findings and some open issues are summarized in \S\,\ref{sec6}.
\section{Problem statement}\label{sec2}

\begin{figure}
\vspace{7mm}
\centering
  \includegraphics[width=0.6\textwidth]{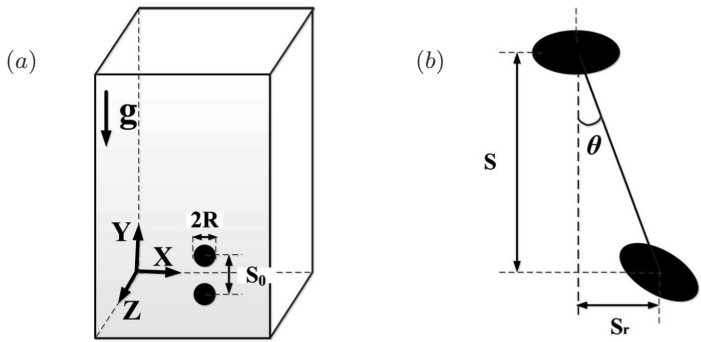}
  
  \vspace{-40mm}
  \hspace{-45mm}$(a)$\hspace{50mm}$(b)$\\
    \vspace{40mm}
  \vspace{2mm}
\caption{Sketch of the problem. $(a)$ initial configuration; $(b)$ definition of some geometric parameters used to characterize the relative position of the two bubbles.}\label{f2.1}
\end{figure}

We consider a pair of deformable gas bubbles rising in line in a large expanse of liquid. The bubbles are assumed to have the same volume $\mathcal{V}$, hence the same equivalent radius $R=\left(3\mathcal{V}/4\pi\right)^{1/3}$. Initially spherical, they are released from rest near the bottom of the numerical tank, with their line of centres vertical ($Y-$ direction) and their centres separated by a distance $S_0$. The initial configuration is illustrated in figure \ref{f2.1}$(a)$. The three-dimensional computational domain is cubic, with a size of $(240R)^3$, which makes it large enough for minimizing artificial confinement effects. The two bubbles start to rise simultaneously under the effect of buoyancy, which is somewhat different from experimental studies in which they are usually released in sequence. During the rise, deviations of the line of centres of the bubble pair from the vertical is characterized by the angle $\theta(t)$ (figure \ref{f2.1}$(b)$).

The liquid and bubble motions are governed by the incompressible one-fluid Navier-Stokes equations
\begin{equation}
\label{e:n-s}
 \nabla\cdot \textbf{\textit{u}}= 0\,,\quad\rho(\partial_t \textbf{\textit{u}}+\textbf{\textit{u}}\cdot \nabla\textbf{\textit{u}})=(\rho-\overline{\rho}) \textbf{\textit{g}}-\nabla p
+ \nabla\cdot\boldsymbol{\Sigma}+\textbf{\textit{F}}_{s}\,.
\end{equation}
In (\ref{e:n-s}), $\textbf{\textit{F}}_{s}=\gamma\kappa\delta_{s}\textbf{\textit{n}}$ stands for the capillary force, with $\gamma$ the surface tension, $\textbf{\textit{n}}$ the unit normal to the interface, $\kappa=-\nabla\cdot\textbf{\textit{n}}$ the interface mean curvature,  and $\delta_s$ the Dirac function identifying the interface position. The suspending liquid and the gas within the bubbles being assumed Newtonian, the viscous stress tensor reads $\boldsymbol{\Sigma}=\mu(\nabla\textbf{\textit{u}}+\nabla\textbf{\textit{u}}^{T})$, with $\mu$ the dynamic viscosity and the superscript ${T}$ standing for the transpose operator. The density $\rho$ and viscosity $\mu$ are uniform in both the liquid and the gas and experience a jump at the interface. A free-slip condition is imposed on all four lateral boundaries, while a periodic condition is assumed to hold on the top and bottom boundaries. In order to ensure that the net momentum flux through the bottom and top planes is constant, and to prevent gravity from accelerating the flow in the vertical direction, the gravity force $\rho \textbf{\textit{g}}$ is supplemented by a body force $-\overline\rho \textbf{\textit{g}}$ with $\overline\rho=f\rho_g+(1-f)\rho_l$, $f$ denoting the global volume fraction of gas in the computational domain, and $\rho_l$ and $\rho_g$ the liquid and gas densities, respectively \citep{bunner2002dynamics}.

In addition to the gas/liquid density and viscosity ratios, usually very small, the dynamics of the system is governed by three independent dimensionless numbers, among which the dimensionless initial separation $\overline{S}_0=S_0/R$. The other two control parameters may be chosen among those listed in table \ref{t2.1}. The Reynolds ($Re$) and Weber ($We$) numbers are generally preferred in theoretical studies. In contrast, in experiments and computations like those discussed here, the terminal rise speed $u_T$ is unknown \textit{a priori}. This is why the Galilei (or Archimedes) number ($Ga$) and the Bond number ($Bo$) are usually selected and are used throughout the present study. The Morton number ($Mo$) is frequently used in place of the Bond number, since $Mo=Bo^3/Ga^4$. In what follows, we vary $Ga$ and $Bo$ in the range $10<Ga<30$ and $0.02<Bo<2.0$, respectively. In this parameter range, an isolated bubble follows a rectilinear path, \textit{i.e.} path instability which is commonly observed for millimeter-size bubbles rising in water does not take place. For such an isolated bubble, selecting $Ga$ and $Bo$ in the above range leads to terminal Reynolds numbers in the range $10\lesssim Re\lesssim120$, depending on bubble deformation. In most of the present work, the initial separation between the two bubble centres is set to $\overline{S}_0=8$, hence the initial gap between the two bubbles is $6R$. In what follows, all variables are normalized using $R$ and $\sqrt{R/g}$ as characteristic length and time scales, respectively. 
The bubble deformation will often be characterized using the aspect ratio $\chi=b/a$, where $b$ and $a$ denote the length of the major and minor axes, respectively. Note that in most available studies, especially the computational works of \cite{Yuan1994} and \cite{hallez2011interaction}, the Reynolds number is based on the bubble diameter rather than the radius. When used for comparison, the corresponding results were converted accordingly.

\begin{table}
\begin{center}
  \begin{tabular}{p{6cm}p{3cm}p{3cm}}
      Name & Abbreviation & Expression    \\[3pt]
       Reynolds number & $Re$ & $\rho_lu_TR/\mu_l$ \\
       Weber number   & $We$ & $\rho_lu_T^2R/\gamma$  \\
       Galilei number   & $Ga$ & $\rho_l g^{1/2}R^{3/2}/\mu_l$  \\
       Bond number   & $Bo$ & $\rho_l gR^2/\gamma$  \\
       Morton number   & $Mo$ & $g\mu_l^4/\rho_l\gamma^3$  \\
  \end{tabular}
  \caption{Dimensionless parameters characterizing the system. $u_T$ is the terminal rise speed of the bubble, and the subscript $l$ refers to the properties of the liquid.}
  \label{t2.1}
  \end{center}
  \vspace{-1mm}
\end{table}

\section{Numerical approach}\label{sec3}
\subsection{\color{black} General features \color{black}}
\label{numer}
\begin{figure}
\centering
 \vspace{2mm}
  \includegraphics[width=0.90\textwidth]{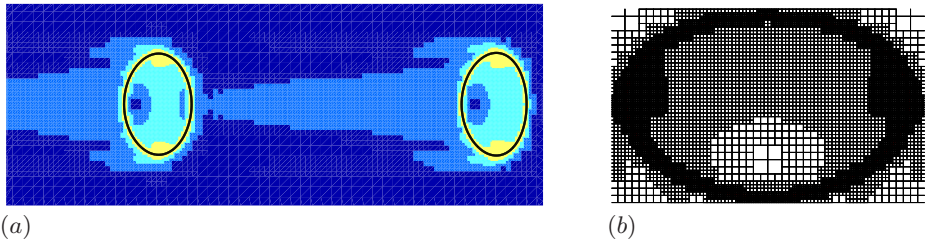}
  \vspace{0mm}\\
  \hspace{-40mm}$(a)$
  \hspace{75mm}$(b)$
\caption{\color{black} Detail of a typical grid for a bubble pair with $Ga=30$ and $Bo=0.3$ at $t=8$. $(a)$: grid refinement in the wake and boundary layer regions; four refinement levels are shown, from dark blue ($\Delta=R/8.5$) to yellow ($\Delta=\Delta_{min}=R/68$). $(b)$: grid detail within the leading bubble and in the neighbourhood of its surface. The gas-liquid interface is marked with a black (resp. red) line in $(a)$ (resp. $(b)$). }
\vspace{-2mm}
\label{gridd}
\end{figure}

The results to be discussed below are obtained by solving (\ref{e:n-s}) with the open source flow solver \textit{Basilisk} developed by Popinet \citep{popinet2009accurate,popinet2015quadtree}. \textit{Basilisk} \color{black} (see basilisk.fr) is the successor of  \textit{Gerris} (http://gfs.sourceforge.net) which has been widely employed over the last fifteen years in detailed explorations of interfacial flows. \color{black} It makes use of Cartesian grids  with a collocated discretization of the velocity and scalar fields. The temporal discretization is based on a second-order fractional step method. In particular, the Godunov-type unsplitted upwind scheme developed by \cite{bell1989second} is used to discretize the advection term, and a fully implicit scheme is used to compute the viscous term. A second-order projection method is employed to ensure that the computed velocity field is divergence-free at the end of each time step. Interfaces are tracked and geometrically reconstructed by a VOF approach in which an accurate well-balanced height function method is used to calculate the interface curvature \color{black} \citep{popinet2009accurate,Popinet2018}. \color{black} An AMR technique allows to locally refine the grid close to interfaces and high-vorticity regions, based on a wavelet decomposition of the gas volume fraction and velocity fields, respectively \citep{van2018towards}. This strategy greatly enhances the computational efficiency while guaranteeing a high numerical accuracy in flow regions where subtle physical phenomena are likely to take place. In the present study, the spatial resolution is refined down to $\Delta_{min} = R/68$ close to the bubble interface. Hence, at the highest Reynolds number considered here ($Re\approx120$), approximately $6$ grid points lie within the boundary layer whose thickness is estimated to $\delta_b \sim Re^{-1/2}$. 
\color{black} Figure \ref{gridd} shows how a typical grid is refined in the vicinity of the two bubbles. In addition to the near-interface zones where $\Delta=\Delta_{min}=R/68$, refined regions include the boundary layer and wake of each bubble, where the local cell size is $\Delta=R/17$. The grid coarsens drastically beyond the region displayed in the figure, the largest cells in the far field corresponding to $\Delta\approx7.5R$. Hence the ratio between the largest and smallest cells in the whole domain is $2^9=512$.\color{black}\vspace{0mm}\\
\indent Previous works have established the capability of \textit{Basilisk} to accurately simulate the dynamics of isolated rising bubbles. \color{black} Moreover, since \textit{Basilisk} succeeded \textit{Gerris} and essentially makes use of the same algorithms, the numerous validations performed with \textit{Gerris} over the years also hold for \textit{Basilisk}. For instance, \cite{web} reproduced successfully with \textit{Basilisk} the results obtained with \textit{Gerris} \color{black} by \cite{cano2016paths} concerning the transition from rectilinear to zigzagging or spiralling paths of air bubbles rising in various liquids. 
In \S\,\ref{sec4.1} we shall compare present predictions for the equilibrium distance of two nearly-spherical bubbles ($Bo\ll1$) rising in line with the findings of \cite{Yuan1994} and \cite{hallez2011interaction}. This comparison will establish the relevance and accuracy of the present approach \color{black} in the axisymmetric case. Nevertheless, this configuration will be proved to be unstable in \S\,\ref{3D}. Therefore, it is necessary to determine to which extent the fate of the three-dimensional system depends on numerical details driving the onset of this instability. This is the purpose of appendix \ref{anum} in which several specific tests are reported.  \\
\indent Most of the computations were run on a personal computer with 24 Intel$^{\mbox{\scriptsize{\textregistered}}}$ Xeon$^{\mbox{\scriptsize{\textregistered}}}$ E5-2630 v3 processors. The Intel-MPI library was used to exchange informations between the processors. A typical run extending over $50$ time units took approximately $50$ days (\textit{e.g.} the case $Ga=30,Bo=0.3$ discussed in \S\,\ref{sec4.4}). Note that, since capillary effects impose a specific time step constraint, low-$Bo$ cases require longer computational times. For instance, to reach a given physical time, the case $Ga=30, Bo=0.05$ was $1.5$ more time consuming than the previous case. Due to the AMR procedure, the grid evolves over time. For $Ga=30,Bo=0.3$, its size stabilizes at approximately $2.5$ million cells during the second half of the run. When additional levels of grid refinement are introduced because the bubbles get very close to each other (see below), this size increases significantly. In such cases, \textit{e.g.} $Ga=20,Bo=0.5$ discussed in \S\,\ref{sec4.5}, the complete grid involves up to $4.2$ million cells.   \color{black}

\subsection{\color{black} Treatment of thin films: numerical vs. physical coalescence}
\label{coalnum}
\begin{figure}
\centering
 \vspace{2mm}
  \includegraphics[width=0.90\textwidth]{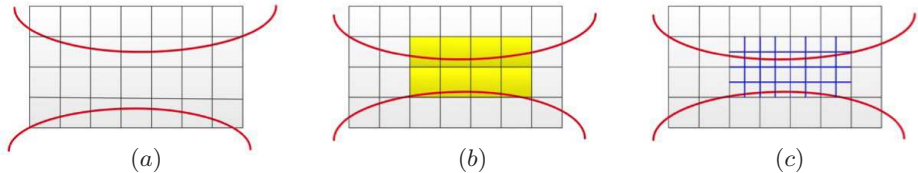}
  \vspace{-2mm}\\
  \hspace{1mm}$(a)$
  \hspace{38mm}$(b)$\hspace{38mm}$(c)$
\caption{Topology-based AMR strategy employed to refine the grid from time $t=t_1$ in $(a)$ to $t=t_1+\Delta t$ in $(b)$ and $(c)$, as the liquid film separating the two bubbles gets squeezed.}
\vspace{-2mm}
\label{f3.1.1}
\end{figure}

\indent In the present problem it is likely that under certain conditions the two bubbles get very close to each other.  \color{black} When this happens in a real flow, coalescence may or may not take place, depending on the mobility of the interfaces involved and on the strength and duration of the forces that drive the two bubbles toward each other \citep{Chesters1991, Chan2011}. Dealing numerically with such situations is particularly challenging, owing to the very small scales involved. \color{black} Some numerical approaches, especially the Front Tracking technique (\cite{unverdi1992,tryggvason2001}), totally prevent coalescence. \color{black} The same may be achieved in VOF approaches by identifying the two bubbles with separate markers, each of them representing the local volume fraction of the corresponding body. 
However, this numerical option is not fully appropriate here. Indeed, for the reasons mentioned below, bubbles rising in line in a pure liquid offer one of the physical situations with the highest coalescence probability.  If coalescence takes place under real conditions, it is of course desirable to track numerically the post-coalescence dynamics, \textit{i.e.} the shape and path evolution of the resulting bubble, which the above option would not allow. \\
\indent That bubbles rising in line in a pure liquid are prone to coalesce in a number of cases is due to the combination of two factors. First, as will be discussed in the next section, the wake of the leading bubble provides a permanent attractive force to the trailing bubble. Under a number of flow conditions, this force is strong enough to make the two bubbles come virtually in contact in the head-on configuration. Second, the mobility of the gas-liquid interfaces when the bubbles are free of any contamination makes the drainage of the interstitial film several orders of magnitude faster than that of liquid-liquid or contaminated gas-liquid interfaces \citep{Vakarelski2018}. 
 The coalescence process may take several distinct forms, depending on the fluid characteristics and bubble size. After the two bubbles collide, the interstitial film may rupture quickly, or the bubbles may stay almost in contact during a long time before eventually coalescing. If viscous effects are small enough, bubbles larger than a critical size bounce after the film has been drained partially and coalesce only after one or several bounces. Energetic considerations and detailed experimental data may be employed to determine which of these scenarios takes place. This issue, together with the specificities of the coalescence of `clean' bubbles are discussed in appendix \ref{coal}.\\
\indent In situations potentially leading to coalescence, one would ideally like to track numerically all steps of the drainage, until non-hydrodynamic effects such as the London-van der Waals force come into play and rupture the film. This typically occurs when the minimum gap between the two interfaces is of the order of $10$ nm. For millimeter-size bubbles, this would require approximately ten additional grid levels beyond the one corresponding to $\Delta_{min}$. Since the largest scales to be resolved in the present context are of the order of $10\,$cm, \textit{i.e.} seven orders of magnitude larger, the corresponding computational cost would be prohibitive.
At least, it is possible to track the first stages of the drainage based on a suitable grid refinement technique. Then, referring to the available knowledge summarized in appendix \ref{coal}, one can reasonably predict which coalescence scenario is taking place in the real system, although the grid resolution does not allow all its details to be captured. The main shortcoming of this approach is that bubbles usually 
merge too early in the computations, \textit{i.e.} they would merge at a higher vertical position in a real flow. Nevertheless, as we discuss in appendix \ref{coal}, the theoretical and experimental knowledge available on the coalescence of nearly-spherical clean bubbles allows the corresponding temporal shift to be estimated in a number of cases. 
\color{black}\\
\indent To track the first steps of the drainage, we developed a specific topology-based AMR scheme to refine the grid within the gap \citep{zhang2019vortex}. The corresponding algorithm checks whether or not any cell crossed by the gas-liquid interface has at least one neighbouring cell filled with only liquid or gas. If not, the cell crossed by the interface is automatically refined. This adaptive strategy is illustrated in figure \ref{f3.1.1}. At $t=t_1$ (panel $(a)$), the two interfaces are separated by a `pure' liquid cell, but this is no longer the case in $(b)$ where the cells standing in the yellow region all contain a nonzero gas fraction. Therefore, these cells are refined by a factor of 2 in each direction, as shown in $(c)$. Ultimately, for the reasons discussed above, we let the two interfaces merge numerically if the number $N$ of successive refinements required to satisfy the above `pure liquid neighbouring cell' criteria exceeds a prescribed value. \color{black} In practice, this is achieved by merging the two markers which identify each bubble in all previous steps. \color{black} The three-dimensional results to be discussed later were all obtained with $N=2$, so that the minimum cell size in the gap was $\delta_{min}=\Delta_{min}/2^N=R/272$. \color{black} For a $0.25\,$mm-radius bubble, this implies $\delta_{min}\approx1\,\mu$m, which is still one hundred times larger than the typical thickness at which the interstitial film actually ruptures.
\color{black} 
\section{Fundamental mechanisms: role of fluid inertia, wake effects and bubble deformation}\label{sec5.2}
\label{mechanisms}
Before analyzing the computational results, a brief review of the main physical mechanisms involved in the problem is in order. First of all, it is key to keep in mind that finite-Reynolds-number interactions between two bubbles are controlled by two antagonistic effects. The first of them is due to the outer flow past the bubble pair, in which irrotational mechanisms prevail. In the potential flow approximation, exact results for the fluid kinetic energy have been established for two spherical bubbles having identical radii, from which the interaction force may be obtained \citep{Voinov1969,Voinov1973,Wijngaarden1976, Miloh1977,Bentwich1978, Biesheuvel1982,Kok1993}. These predictions indicate that the interaction is repulsive when the two bubbles rise in line because the fluid velocity reaches a minimum in the gap, inducing a pressure maximum there \citep{Harper1970}. Conversely, the interaction is attractive when the bubbles rise side by side, owing to the flow acceleration (hence the pressure minimum) in the gap. The critical angle at which the interaction force changes sign depends on the separation between the two bubbles, ranging from $35^\circ$ when they are in contact to a value close to $54^\circ$ when they are far away from each other \citep{Kok1993}.\\
\indent Finite-Reynolds-number effects manifest themselves in the generation of vorticity at the bubble surface, owing to the shear-free condition obeyed by the carrying liquid when the gas-to-liquid viscosity ratio is negligibly small and the interface is uncontaminated by surfactants. Diffusion and advection of this vorticity in the surrounding fluid results in a boundary layer and a wake past each bubble. Vortical effects at the bubble surface lower the pressure at the rear stagnation point compared to that at the front \citep{Kang1988}, so that the pressure at a given position along the wake axis is lower than it would be in the potential flow limit. When the bubbles rise in line, this pressure drop makes the trailing bubble (hereinafter abbreviated as TB) sucked toward the leading bubble (hereinafter abbreviated as LB). In contrast, when they rise side by side, the interaction of the two wakes results in a pressure maximum in the gap, hence a force tending to move the two bubbles away from each other.\\
\indent In a given geometrical configuration of the tandem, the relative magnitude of the above two antagonistic effects depends on the Reynolds number, assuming the bubbles to keep a spherical shape. Vortical effects dominate when $Re$ is low enough, while the interaction is expected to become close to potential flow predictions for large enough $Re$. For this reason, keeping the Reynolds number and the inclination of the line of centres fixed, the overall interaction force vanishes when the separation between the two bubbles takes a specific value, $\overline{S}_e(Re,\theta)$. The smaller $Re$ is, the shorter (resp. larger) this equilibrium separation is for $\theta=0^\circ$ (resp. $\theta=90^\circ$). The two bubbles collide if $\overline{S}_e$ is small enough, \textit{i.e} $\overline{S}_e\leq2$ for spherical bubbles. Then, assuming that the interface is uncontaminated, they may either coalesce, bounce 
or stay in contact for a very long time, 
depending on whether or not the net attractive force is large enough to achieve the drainage of the interstitial film. 
Approximate models have been proposed to predict $\overline{S}_e$ in the in-line configuration for spherical bubbles. These models consider that the wake of the LB, taken into account by using Oseen or high-$Re$ far-wake velocity distributions, decreases the fluid vertical velocity `felt' by the TB at a given position by an amount equal to the cross-sectional average of the velocity defect at that position \citep{katz1996wake,ramirez2011drag,ramirez2013forces}. \\
\indent 
When the TB moves behind the LB with some offset from the wake axis it faces a non-uniform, asymmetric flow which may locally be considered as a shear flow. A spherical bubble rising in a linear shear flow is known to experience a shear-induced lift force \citep{auton1987lift,legendre1998lift}. If the bubble moves faster than the fluid along the streamlines of the base flow, this sideways force deviates it toward the direction of the descending fluid. In the in-line configuration, the relative flow faced by the TB moves downwards and its velocity grows with the distance to the wake axis. Therefore the sideways force tends to move the TB out of the wake of the LB, making the in-line configuration unstable with respect to an infinitesimal lateral deviation \citep{Harper1970}. As will be seen later, this instability plays a crucial role in the evolution of a bubble pair.\\
\indent The mechanisms reviewed so far subsist of course when bubble deformation becomes significant. However their magnitude is deeply influenced by the bubble shape. In particular, a key feature of vorticity generation on a curved shear-free interface is that the resulting tangential vorticity, say $\omega_s$, is proportional to the product of the local surface curvature and tangential velocity of the fluid \citep{Batchelor1967}. This makes the magnitude of $\omega_s$ increase with bubble deformation. In contrast, for a given interface shape, $\omega_s$ does not depend on $Re$ when the Reynolds number is large, unlike the more familiar case of a no-slip surface. In the limit $Re\gg1$, the maximum of $\omega_s$ (normalized by the rise velocity and equivalent bubble radius) at the surface of an oblate bubble with an aspect ratio $\chi$ increases by a factor of $4$ from $\chi=1$ (spherical bubble) to $\chi=2$, eventually growing as $\chi^{8/3}$ when $\chi\gg1$ \citep{magnaudet2007wake}. In the in-line configuration, this dramatic increase implies that the attraction of the TB toward the LB becomes increasingly strong as the latter deforms. Consequently, bubble deformation is expected to reduce significantly the equilibrium separation, favouring coalescence. Another consequence of deformation is its influence on the magnitude and even the sign of the sideways force acting on the TB when the axial symmetry of the in-line configuration is broken by some lateral disturbance. The corresponding mechanisms are discussed in appendix \ref{vortrev}.
\section{ Axisymmetric configuration}\label{sec4.1}
\begin{figure}
 \vspace{5mm}
  \centerline{\includegraphics[width=0.8\textwidth]{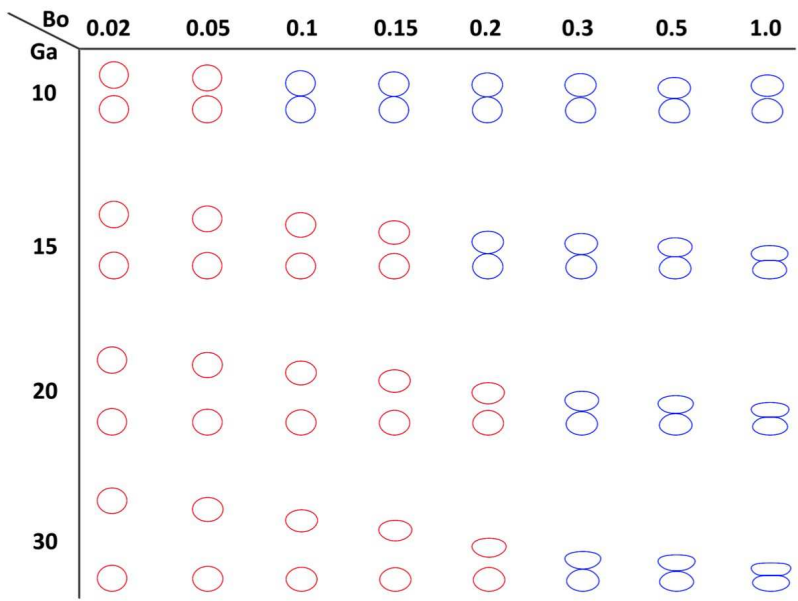}}
  \vspace{3mm}
  \caption{Final bubble shapes and separations observed in the axisymmetric configuration. Red bubbles maintain a finite separation, while blue bubbles are about to coalesce.}
\label{f4.1.1}
\end{figure}

Numerical simulations of the in-line configuration based on boundary-fitted grids were reported by \cite{Yuan1994} and \cite{hallez2011interaction} for spherical bubbles. Here in contrast, the bubbles deform freely and continuously while rising, which allows us to investigate the influence of their deformation on the interaction process. In this section, we restrict the generality of the problem by constraining the bubbles to follow a straight vertical path. For this purpose, instead of the cubic domain described earlier, we use an axisymmetric domain whose radius and height are both $240R$. Bubbles are released on the symmetry axis are rise along it. Computations are carried out with gas/liquid density and viscosity ratios set to $10^{-3}$ and $10^{-2}$, respectively. 
\\
\indent Figure \ref{f4.1.1} displays the final bubble shapes and relative positions obtained through 32 computational runs covering the domain ($10\leq Ga\leq30$, $0.02\leq Bo\leq1.0$), all with $\overline{S}_0=8$. Red bubbles maintain an equilibrium distance which is finite in the computational sense, \textit{i.e.} the gap that separates them at steady state exceeds the minimum cell size $\delta_{min}$ allowed by the specific topology-based AMR treatment described in \S\,\ref{coalnum}. Conversely, blue bubbles are such that the `final' gap is thinner than $\delta_{min}$, implying that coalescence is about to take place numerically.
Consider a given row in the figure, \textit{i.e.} a given $Ga$. Increasing $Bo$ increases the ability of the bubbles to deform. Hence it reduces the final equilibrium separation distance $\overline{S}_e$ between their centroids, which makes coalescence more likely to occur.
For $Ga=10$, increasing the Bond number from $Bo=0.02$ to $Bo=0.1$ makes the Weber number increase from $We=0.18$ to $We=0.90$. As a consequence, the final deformation of the LB and that of the TB rise from $\chi=1.02$ to $\chi=1.1$ and from $\chi=1.01$ to $\chi=1.04$, respectively. Although still modest, this deformation makes the bottom region of the LB significantly flatter than the top region of the TB, allowing a thin-gap region with a finite area to develop. However, this deformation is small enough for the rise speed to remain virtually unaffected, which leaves the terminal Reynolds number unchanged throughout this range of $Bo$ ($Re\approx21$).  
That the LB deforms more than the TB becomes clearer as $Ga$ increases. This is directly due to the fact that the wake of the former reduces the pressure at the front stagnation point of the latter, a mechanism often referred to as the `sheltering' effect. Hence the pressure difference between the front stagnation point and the bubble equatorial plane, which drives the deformation \citep{moore1959,moore1965}, is smaller on the TB. For $Ga=30$, the final separation distance decreases from $\overline{S}_e=5.72$ to $\overline{S}_e=3.63$ when the Bond number increases from $0.05$ to $0.15$. The Weber number is now of $\mathcal{O}(1)$, increasing from $0.65$ to $1.44$. At the same time, the Reynolds number decreases from $108$ to $91$, due to the drag increase associated with the increasing oblateness of the two bubbles. 
\begin{figure}
\vspace{3mm}
  \centerline{\includegraphics[width=0.96\textwidth]{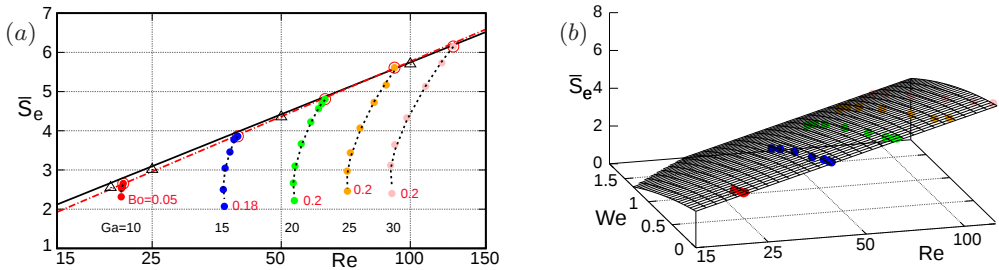}}
  \vspace{-33.5mm}
  \hspace{1mm}$(a)$\hspace{69mm}$(b)$
   \vspace{31mm}
  \caption{Variation of the final equilibrium distance $\overline{S}_e$ against: $(a)$ the Reynolds number; and $(b)$ the Reynolds and Weber numbers. Both $Re$ and $We$ are based on the final rise velocity of the bubble pair. Each series identified with a given color corresponds to the same $Ga$ and different $Bo$ (increasing from top to bottom in $(a)$ and from front to back in $(b)$). In $(a)$, the open circles correspond to $Bo=0.005$, while the solid line and open triangles refer to the prediction (\ref{e4.1}) and the numerical results of \cite{hallez2011interaction} for spherical bubbles, respectively. The dotted lines in ($a)$ and the curved surface in $(b)$ correspond to the prediction (\ref{e4.2}), while the dash-dotted line in $(a)$ represents this prediction  evaluated for $We=0$. }
\label{f4.1.2}
   \vspace{-2mm}
\end{figure}

The variation of $\overline{S}_e$ with $Re$ and $We$ may be obtained by considering the bubble pairs of figure \ref{f4.1.1} once they have reached their final configuration. The result is displayed in figure \ref{f4.1.2}. 
Specific runs were carried out with $Bo=0.005$ (hollow circles in figure \ref{f4.1.2}$(a)$) to maintain the interface shape very close to a sphere, in order to compare present predictions with results available for spherical bubbles. For $(Ga, Bo)=(30,0.005)$, the final aspect ratio of the LB (TB) is $\chi=1.023$ ($\chi=1.017$), and the deformation is even less for lower $Ga$. Therefore, the corresponding equilibrium distance is expected to agree well with the correlation proposed by \cite{Yuan1994}, namely
\begin{eqnarray}
\label{e4.1}
\overline{S}_e(Re) &=& 4.40~\textrm{log}_{10}~Re-3.06
\end{eqnarray}

\noindent As figure~\ref{f4.1.2}$(a)$ reveals, present low-$Bo$ predictions are in excellent agreement with (\ref{e4.1}) for $Ga\geq20$. For lower $Ga$, the numerical values of $\overline{S}_e(Ga,Bo=0.005)$ are found to lie slightly below the prediction (\ref{e4.1}). This is in line with the results of \cite{hallez2011interaction} which indicate for instance that $\overline{S}_e(Re=20)$ is approximately $5\%$ less than predicted by (\ref{e4.1}). Figure~\ref{f4.1.2}$(a)$ shows that present results for $Bo=0.005$ (open circles) are in excellent agreement with those of the latter authors (open triangles) throughout the range of Reynolds number considered here. This establishes the accuracy of present computations in the in-line configuration. 

As $Bo$ increases, the equilibrium separation quickly falls below that predicted by (\ref{e4.1}) and a correlation in the form $\overline{S}_e=f(Re,We)$ must be sought. 
A first attempt aimed at recovering (\ref{e4.1}) in the limit $We\rightarrow0$ was not successful. After several trials, we found that the entire set of present results is best approached by the fit
\begin{eqnarray}
\label{e4.2}
\overline{S}_e(Re,We) &=& 2.025\,\textrm{log}\,Re - 3.56- 0.98\,We - 0.36\,We^2\,. 
\end{eqnarray}
Figure \ref{f4.1.2}$(a)$ indicates that, once the Weber number has been properly eliminated ($We=(Re/Ga)^2Bo$), all numerical data collapse onto the corresponding curves (dotted lines). The three-dimensional representation of figure \ref{f4.1.2}$(b)$ confirms that numerical data all fall on the curved surface defined by (\ref{e4.2}). Therefore (\ref{e4.2}) is seen to provide a valid prediction of the equilibrium separation distance at least in the range $20\lesssim Re\lesssim 120$ and $0 < We \lesssim 1.5$. Actually, we also performed some computations for $Ga=40$ and $Ga=50$ and found that the equilibrium separation obtained in the range $0.02\leq Bo\leq0.2$ is still correctly predicted by (\ref{e4.2}). Note that, although (\ref{e4.2}) does not reduce to (\ref{e4.1}) in the limit $We\rightarrow0$, 
the corresponding fit (red dash-dotted line in figure \ref{f4.1.2}$(a)$) achieves a better agreement than (\ref{e4.1}) with present low-$Bo$ predictions, as well as with the results of \cite{hallez2011interaction}. Nevertheless, it must be kept in mind that neither (\ref{e4.1}) nor the low-$We$ limit of (\ref{e4.2}) remains valid for $Re\lesssim15$, since both expressions predict $\overline{S}_e<2$ at lower $Re$. In the low-but-finite Reynolds number range, asymptotic results for rigid spheres \citep{Happel1965} may readily be transposed to bubbles using the scaling argument developed by \cite{Legendre1997}. By doing so, it is concluded that the two bubbles always collide for $Re\lesssim1$, since the LB experiences a larger drag than the TB. Actually, there are computational indications that collision takes place as soon as $Re\leq15.5$ \citep{watanabe2006line}, which corresponds well to the threshold predicted by (\ref{e4.2}) ($\overline{S}_e(Re,We=0)=2$ for $Re=15.55$). 
\section{Three-dimensional configurations}
\label{3D}
\subsection{Overview of the results}\label{sec4.2}

\begin{figure}
\vspace{5mm}
\centering
  \includegraphics[width=0.98\textwidth]{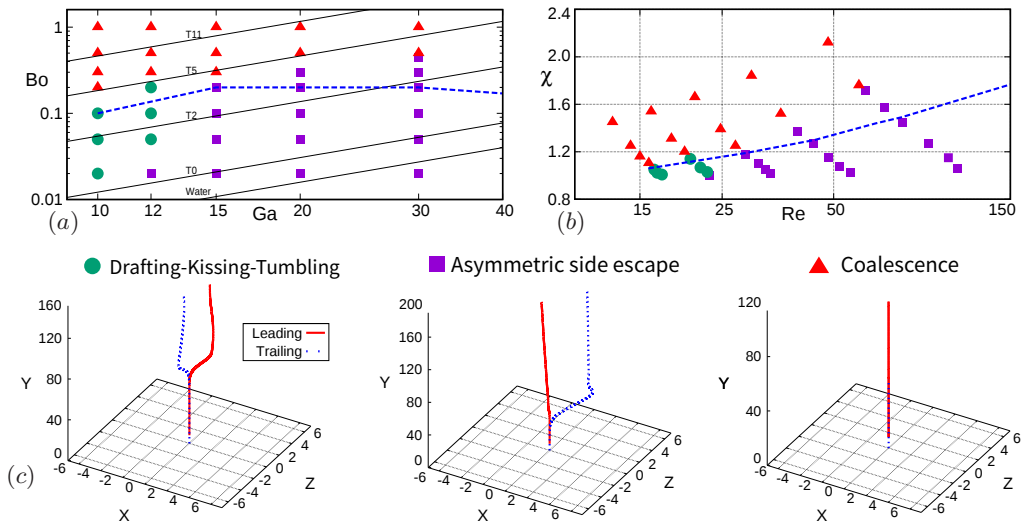}
  \vspace{2mm}
\caption{Set of configurations encountered in the three-dimensional simulations. $(a)$ $(Ga, Bo)$ phase diagram; $(b)$ $(Re, \chi)$ phase diagram (for each $(Ga,Bo)$ pair, $Re$ and $\chi$ are the steady-state values determined with the corresponding isolated bubble); $(c)$ typical trajectories illustrating each of the three configurations. Bullets: DKT scenario; solid squares: ASE scenario; solid triangles: collision followed by coalescence. In $(c)$, the three panels from left to right correspond to $(Ga, Bo) = (10, 0.1)$, $(30, 0.3)$ and $(20, 0.5)$, respectively. The thin solid lines in $(a)$ are the iso-$Mo$ lines corresponding to different liquids, with water at the very bottom, then silicone oils T0-T11 of increasing viscosity from bottom to top (\textit{e.g.} \cite{Zenit2008} for the corresponding physical properties). The dashed line in $(a)$ and $(b)$ separates the subdomain of weakly deformed bubbles in which a finite equilibrium separation is reached in the axisymmetric case, from that in which the bubbles eventually coalesce. }
  \vspace{-84.5mm}
  \hspace{-55mm}$(a)$\hspace{63mm}$(b)$\\
  \vspace{30mm}
   \hspace{-131mm}$(c)$
     \vspace{80mm}
\label{f4.2.1}
  \vspace{-34mm}
\end{figure}
As mentioned earlier, it is known since \cite{Harper1970} that the in-line configuration of two clean spherical bubbles is unstable with respect to side disturbances when the Reynolds number is large. 
Experimental investigations carried out under surfactant-free conditions (silicone oils) in the range $10\lesssim Re\lesssim150$ confirm this prediction \citep{Kusuno2015,Kusuno2019}. We performed fully three-dimensional time-dependent simulations to assess this stability issue and explore its consequences. Similar to the axisymmetric case, the gas/liquid density and viscosity ratios were set to $10^{-3}$ and $10^{-2}$, respectively. In agreement with the above experimental and theoretical findings, we observed that the bubble pair never maintains a straight vertical path except when the Bond number is large enough for coalescence to eventually happen. We actually identified three drastically different evolutions, depending on the value of the Galilei and Bond numbers. Beyond a critical Bond number, $Bo_c(Ga)$ which increases approximately from $0.2$ for $Ga=10$ to $0.5$ for $Ga=30$, the two bubbles collide and eventually coalesce\color{black}, this `coalescence' having to be interpreted in the light of the discussion of \S\,\ref{coalnum} (see below)\color{black}. Predictions of the three-dimensional and axisymmetric simulations superimpose for $Bo\gtrsim0.5$ when $Ga\leq20$ (and for $Bo\gtrsim1$ when $Ga=30$), indicating that the in-line configuration is stable with respect to azimuthal disturbances for sufficiently deformed bubbles. In contrast, for $Bo<Bo_c(Ga)$, the TB escapes from the wake of the LB at some point, and the two go on rising with their line of centres more or less inclined with respect to the vertical and their centroids widely separated. In such cases, we found that the interplay of the two bubbles after the three-dimensional effects set in may follow two markedly different scenarios. One is clearly a DKT-type mechanism. 
 In this case, both bubbles deviate from their initial trajectory and eventually rise almost side by side along two straight vertical lines distinct from the initial path. In contrast, in the other scenario, which we refer to as Asymmetric Side Escape (hereinafter abbreviated as ASE), the lateral drift of the TB leaves the path of the LB almost unaffected. Hence this bubble goes on rising virtually along its initial path, while after the system has reorganized itself, the TB follows a markedly distinct vertical path. The structural differences between the configurations corresponding to the DKT and ASE scenarios are well visible in the recent observations of \cite{Kusuno2019}; see especially their figures 4 and 5.\\
\indent Figure~\ref{f4.2.1} displays the phase diagrams and typical paths corresponding to the above three evolutions, still for $\overline{S}_0=8$. The influence of initial conditions, \textit{i.e.} angular inclination and separation, on the borders of the different subdomains will be discussed in \S\S\,\ref{devi} and \ref{sec5.3}, respectively. \color{black} The influence of numerical parameters, among which the grid resolution, on the coalescence threshold, \textit{i.e.} on $Bo_c(Ga)$, is estimated in appendix \ref{anum}. These technical aspects are found to barely change $Bo_c(Ga)$ by a few percent. 
\begin{figure}
\vspace{5mm}
  \centerline{\includegraphics[width=0.8\textwidth]{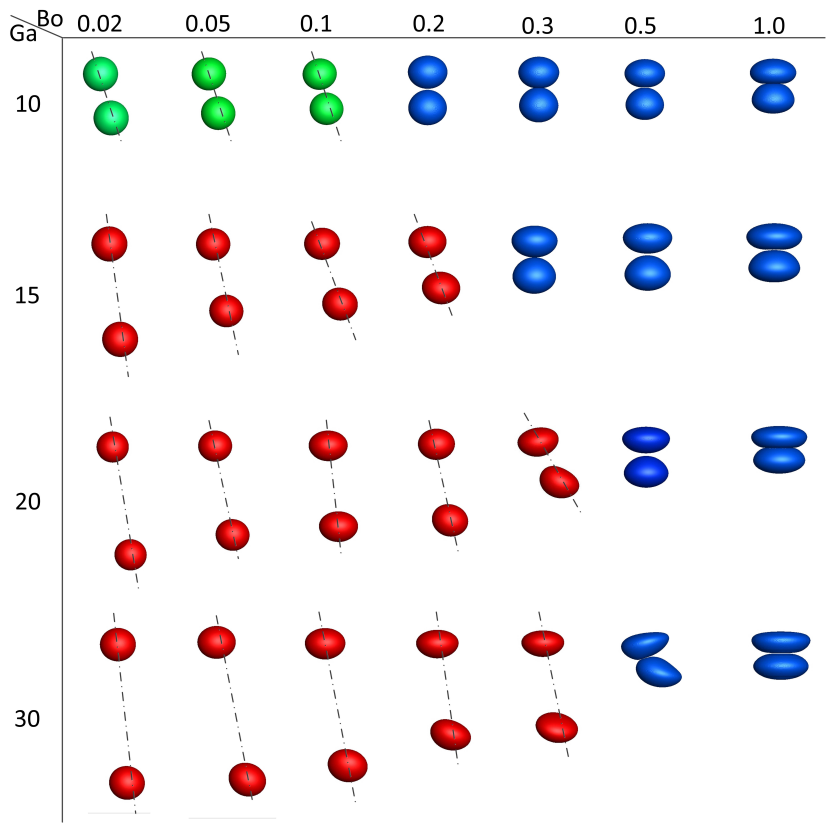}}
  \caption{Snapshots of the bubble shapes and relative positions observed during the lateral escape of the TB or the pre-coalescence process. Red, green and blue pairs correspond to the ASE, DKT, and coalescence scenarios, respectively. In the red and green regions, the snapshots are taken by the time the horizontal distance between the two centroids is approximately equal to one bubble initial radius; in the blue region, each snapshot is the last in a run before numerical coalescence occurs. Since successive snapshots are separated by a finite time interval, the remaining time until coalescence may differ among the various blue pairs.}
  \vspace{-3mm}
\label{f4.2.2}
\end{figure}
As the thin black lines indicate, coalescence only takes place in the present configuration in liquids with a sufficiently high Morton number, from $Mo\approx8\times10^{-7}$ for $Ga=10$ to $Mo\approx1.5\times10^{-7}$ for $Ga=30$. This means in particular that bubbles rising in pure water ($Mo\approx2.6\times10^{-11}$) never coalesce in the $Ga$-range considered here. \color{black} Interestingly, figures~\ref{f4.2.1}$(a)-(b)$ also reveal that under some conditions such as $(Ga,Bo)=(20,0.2)$ or $(30,0.2)$, bubbles that would coalesce if the system were constrained to remain axisymmetric actually escape from coalescence through the ASE scenario. From figure \ref{f4.2.1}$(b)$, it may also be concluded that for $Re\approx15$, even an $8\%$ departure from sphericity is sufficient to lead to coalescence.  \color{black}Actually, the discussion of appendix \ref{coal} indicates that, for such modest Reynolds numbers, viscous effects tremendously delay the drainage of the interstitial film. This is why experiments performed in this regime \citep{Sanada2006,watanabe2006line} reveal that, after the two bubbles collide, they merely stay `glued' to each other and rise as a single `dumbbell' bubble without coalescing during the time window of the observations. However, since all forces involved in the physical system, including the London-van der Waals force, are attractive, there is no doubt that such bubbles eventually coalesce. Unfortunately, the spatial resolution of the present simulations is clearly insufficient to reproduce this slow coalescence process. That is, the computations correctly predict the final state of the physical system but this state is reached too early. This underestimate of the coalescence time subsists at larger Reynolds number (see \S\,\ref{sec4.5}) but, for reasons discussed in appendix \ref{coal}, reduces as $Re$ increases. 
 \color{black} 
In figure \ref{f4.2.1}$(b)$, the maximum deformation below which the two bubbles do not coalesce is seen to increase significantly with $Re$. For instance, bubbles with $\chi\approx1.5$ follow an ASE scenario for $Re\approx70$, but coalesce for $Re\approx35$ (in these estimates, $\chi$ and $Re$ are evaluated from the steady-state properties of the isolated bubble corresponding to the same $(Ga,Bo)$ set).\\
\indent Figure \ref{f4.2.2} depicts the bubble shapes and relative positions during the lateral escape of the TB or the coalescence process.  
Considering the DKT (green) and ASE (red) regimes at a fixed $Ga$, the figure indicates that the larger $Bo$ the shorter the separation during the escape stage. This may be interpreted as a stabilizing effect of the deformation, since the bubble pair is able to maintain a vertical path during a longer time (\textit{i.e.} down to a shorter separation) when the Bond number increases. Then, coalescence takes place when $Bo$ exceeds the critical threshold $Bo_c(Ga)$ which is seen to increase significantly with $Ga$, as already noticed in figure \ref{f4.2.1}$(a)$. In most cases, coalescence is reached through a head-on approach, \textit{i.e.} without any prior escape of the TB, the axisymmetric film in the gap being gradually squeezed. However, for $Ga=30$ and $Bo=0.5$, an intermediate configuration corresponding to an oblique approach is noticed. These different regimes are examined in more detail in the rest of this section. 

\subsection{Drafting-Kissing-Tumbling}\label{sec4.3}

\begin{figure}
\vspace{5mm}
\centering
   \includegraphics[width=0.98\textwidth]{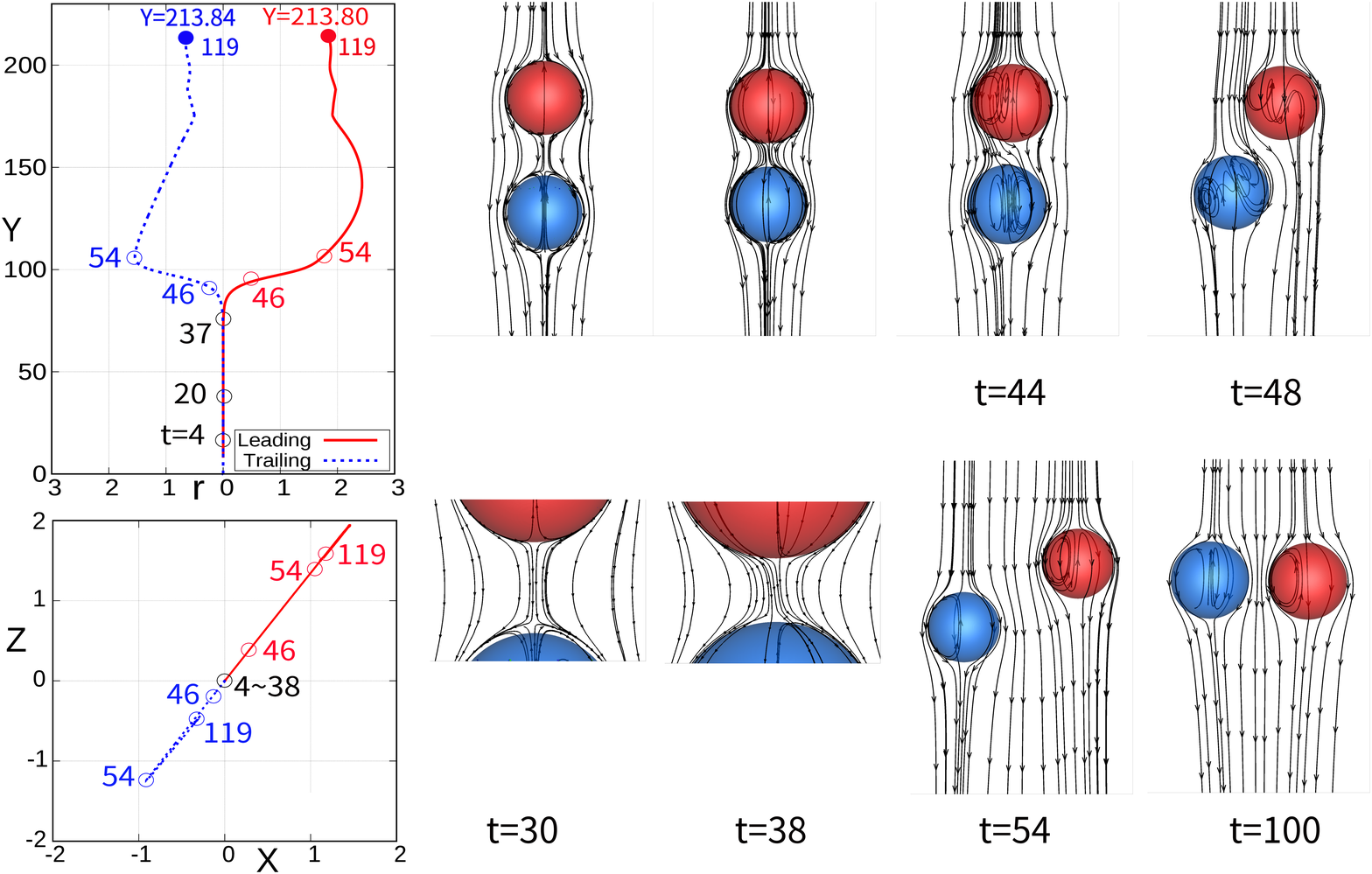}
   \vspace{-84mm}\\
   \hspace{-90mm}$(a)$\hspace{41mm}$(b)$
   \vspace{84mm}
\caption{Path of a bubble pair following a DKT scenario ($Ga=10,\,Bo=0.02$). $(a)$ Side and top views, with $r$ the radial distance to the initial path and numbers referring to the dimensionless time at the corresponding position; $(b)$ three-dimensional streamlines past the bubbles at different instants of time, in the reference frame of the leading bubble. A zoom of the flow field in the gap is provided for $t=30$ and $t=38$, to highlight the inception of the non-axisymmetric fluid motion.}
\label{f4.3.1}
\end{figure}

The DKT scenario followed by sedimenting particle pairs has been widely described for spheres \citep{joseph1986nonlinear, fortes1987nonlinear, Feng1994}, thick disks \citep{brosse2014interaction}, and under certain conditions prolate spheroids \citep{ardekani2016numerical}. In short, owing to the sheltering effect induced by the wake of the leading body, the trailing body first catches up with it (drafting) until the two collide (kissing). Then the resulting prolate compound body becomes unstable to transverse disturbances, which makes it rotate (tumbling) in such a way that the line joining the centroids of the two individual bodies tends to become horizontal, letting them eventually fall/rise separately in a side-by-side configuration. A similar behaviour of bubbles rising in line has been reported experimentally by \cite{Kusuno2019} in the range $10\lesssim Re\lesssim25$, $0.3\lesssim We\lesssim1.1$. Here we identified it for bubble pairs corresponding to $Ga\lesssim12$ and $Bo\lesssim0.2$. With $(Ga, Bo)=(10, 0.02)$, \textit{i.e.} $Mo=8\times10^{-10}$, the final state corresponds to $Re\approx16$ and $\chi\approx1.02$, which indicates that the bubbles shape remains close to a sphere. The corresponding evolution is displayed in figure \ref{f4.3.1}.

\begin{figure}
\centering
  \vspace{28mm}
  \hspace{54mm}$(a)$\hspace{63mm}$(b)$\\
  \vspace{22.3mm}
    \hspace{54mm}$(c)$\hspace{63mm}$(d)$\\
      \vspace{-51mm}
  \includegraphics[width=0.95\textwidth]{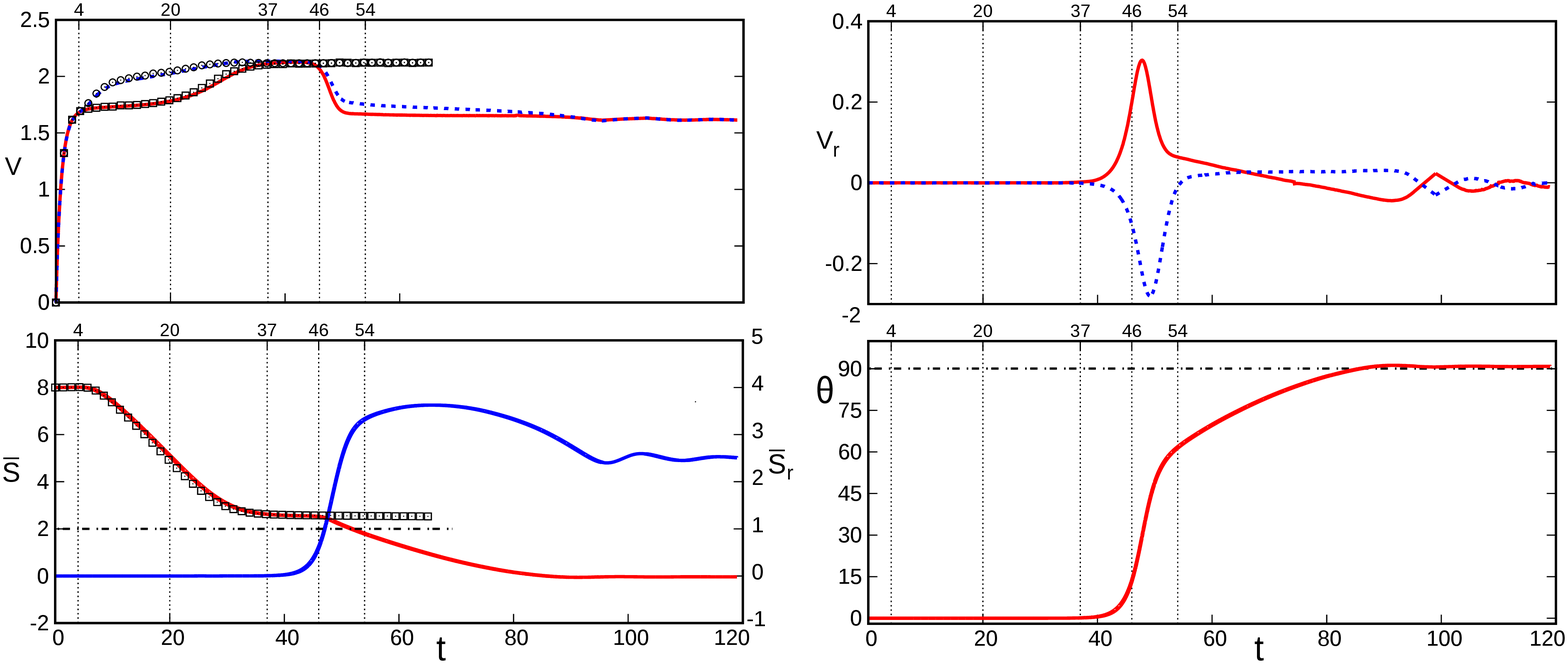}

\caption{Evolution of several characteristics of a bubble pair with $(Ga, Bo)=(10, 0.02)$ undergoing a DKT interaction. $(a)$: vertical velocity component of the LB (solid red line) and TB (dashed blue line); $(b)$ same for the horizontal component; $(c)$: vertical (red line, left axis) and horizontal (blue line, right axis) separations; $(d)$ inclination of the line of centres with respect to the vertical. Open squares and circles in $(a)$ and $(c)$: axisymmetric prediction. 
}
\label{f4.3.2}
\end{figure}

As revealed in panel $(a)$, the tumbling process starts at $t\gtrsim37$ when the two bubbles reach the same rise velocity (see figure \ref{f4.3.2}$(a)$). After this process is completed (say at $t=54$ in figure \ref{f4.3.1}$(a)$), both paths have experienced large lateral deviations. Their lateral separation is about $2.3$, the LB having been slightly more deviated. They stand virtually side by side and keep on rising in this configuration, although the rise velocity of the (formerly) TB is barely larger than that of the LB, inducing a tiny difference in the final altitude reached by the two bubbles. During the side-by-side rise, the lateral separation $\overline{S}_r$ evolves significantly, until an equilibrium value $\overline{S}_{re}\simeq2.55$ is reached. This is consistent with the findings of  \cite{Legendre2003} which indicate that in this configuration, the transverse force acting on a pair of spherical bubbles almost vanishes for  $\overline{S}_r=2.5$ when $Re=25$ (their figure 13). For $Y>175$, the centre of inertia of the bubble tandem is seen to drift slightly towards the left, a consequence of the tiny inclination of its line of centres. The sign of this drift is consistent with the conclusions of \cite{hallez2011interaction}  who found (their figure 9$(b)$) that in the range $10\leq Re\leq 25$, the centre of inertia drifts laterally toward the position of the higher bubble, provided $40^\circ\lesssim\theta<90^\circ$. The horizontal trace of the TB and LB paths in figure \ref{f4.3.1}$(a)$ indicates that the entire motion remains planar and takes place within a vertical plane. Hence, by breaking the initial axial symmetry of the flow, the DKT mechanism changes the initial one-dimensional path into a two-dimensional path, but the latter keeps track of the former since the preferential direction of the bubble motion remains unchanged.


Some three-dimensional streamlines past the two bubbles are displayed in figure \ref{f4.3.1}$(b)$. No standing eddy is observed at the back of the bubbles. This is in line with the generating mechanism of vorticity on a curved shear-free surface discussed in \S\, \ref{mechanisms}, which maintains the flow past a shear-free sphere unseparated whatever the Reynolds number \citep{blanco1995structure,magnaudet2007wake}. 
At $t=38$, the separation has almost reached its minimum ($\overline{S}\approx2.3$) and the flow is still almost axisymmetric, except within the gap where a small left-right asymmetry is discernible in the enlarged view. In the present case, the origin of the axial symmetry breaking stands in tiny numerical asymmetries, \color{black} in the first place those resulting from the pressure field returned by the multilevel Poisson solver (see appendix \ref{anum}). \color{black} Tumbling then starts. At $t=44$ and $t=48$, the flow field exhibits strong left-right asymmetries which result in transverse forces that act to move the two bubbles in opposite directions. Later, say for $t\geq54$, the flow field gradually rearranges towards a left-right symmetric configuration corresponding to a side-by-side motion of the two bubbles, with an equilibrium horizontal separation $\overline{S}_{re}\approx2.55$.\vspace{1mm}\\
\indent The evolution of the rise speed of the two bubbles is plotted in figure \ref{f4.3.2}$(a)$. 
The three-dimensional prediction is seen to coincide with its axisymmetric counterpart up to $t\approx45$, \textit{i.e.} throughout the time period during which the bubble pair moves in straight line and even during the early stage of the tumbling process. That the three-dimensional and axisymmetric predictions virtually superimpose up to $t=37$ proves the reliability of the former. Note that the red and blue curves superimpose up to $t=4$, which corresponds to the very early stage during which the two bubbles rise independently. During the next stage, say $4<t<20$, the TB goes on accelerating while the rise speed of the LB (hereinafter denoted as $V_{LB}$) stays almost constant. This corresponds to the early stage of the interaction during which the trailing bubble is sucked in the wake of the LB while the latter remains unaffected. Then, from $t=20$ to $t=37$, the two bubbles get close enough for the TB to modify the wake of the LB, the rise speed of which increases sharply until the two bubbles rise with the same speed. 
  Tumbling starts at $t\approx38$, without any real `kiss' since figure \ref{f4.3.2}$(c)$ indicates that $\overline{S}$ never fell below $2.3$ at previous times. Up to $t=46$, no change is noticed in the rise speed of the two bubbles, nor in their vertical separation, although the tumbling process is going on, as the sharp rise of their lateral velocities (figure \ref{f4.3.2}$(b)$) and that of the inclination of their line of centres (figure \ref{f4.3.2}$(d)$) confirm. 
 The situation significantly changes within the next short stage $46\leq t\leq50$, during which $V_{LB}$ and  $V_{TB}$ (the rise speed of the TB) drop sharply and the line of centres rotates by more than $30^\circ$. Not surprisingly, this rotation forces the lateral separation to increase dramatically, from $\overline{S}_{r}\approx1$ at $t=46$ to $\overline{S}_{r}\approx2.5$ at $t=50$. Conversely, the vertical separation is reduced to $\overline{S}\approx2$ at $t=50$. The tumbling motion is also responsible for the slowing down of the two bubbles, as it makes the flow around them fully three-dimensional, which enhances the dissipation in the liquid, hence the drag on each bubble. 
 Tumbling goes on more slowly until $t\approx85$, when the side-by-side configuration ($\theta=90^\circ$) and the equilibrium horizontal separation ($\overline{S}_{re}\approx2.55$) are reached. Ultimately, $\theta$ slightly exceeds $90^\circ$, in line with the tiny difference between the final positions of the two bubbles noticed in figure \ref{f4.3.1}$(a)$.


\subsection{Asymmetric Side Escape}\label{sec4.4}

\begin{figure}
\vspace{5mm}
\centering
  \includegraphics[width=0.96\textwidth]{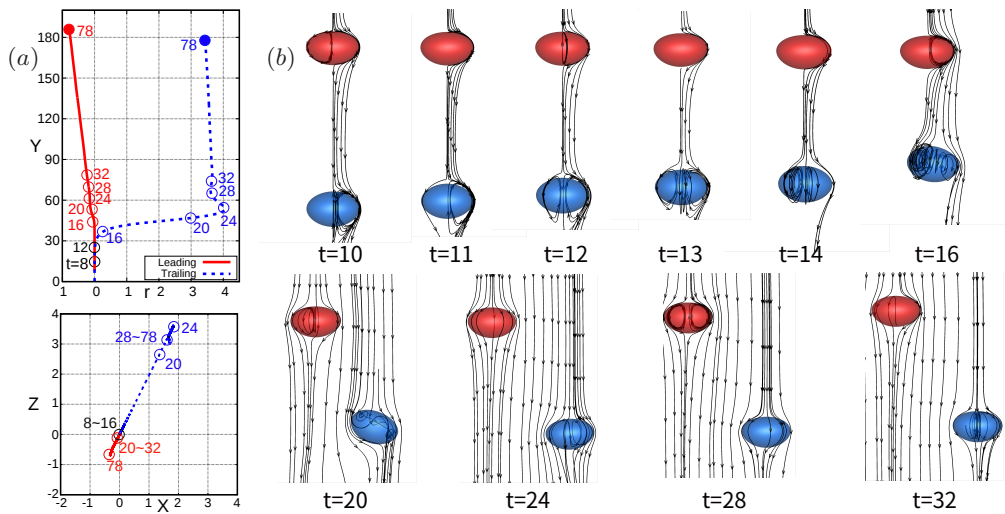}
   \vspace{-62.5mm}\\
   \hspace{-97mm}$(a)$\hspace{30mm}$(b)$
   \vspace{60.5mm}
\caption{Path of a bubble pair following an ASE scenario ($Ga=30,\,Bo=0.3$). $(a)$ Side and top views of the path with numbers referring to the dimensionless time at the corresponding position of the TB; $(b)$ three-dimensional streamlines past the bubbles at different instants of time, in the reference frame of the leading bubble. }
\label{f4.4.1}
\vspace{-2mm}
\end{figure}
\begin{figure}
\vspace{-14mm}
\centering
 \vspace{38mm}
  \hspace{54mm}$(a)$\hspace{63mm}$(b)$\\
  \vspace{22.3mm}
    \hspace{54mm}$(c)$\hspace{63mm}$(d)$\\
      \vspace{-50mm}
  \includegraphics[width=0.95\textwidth]{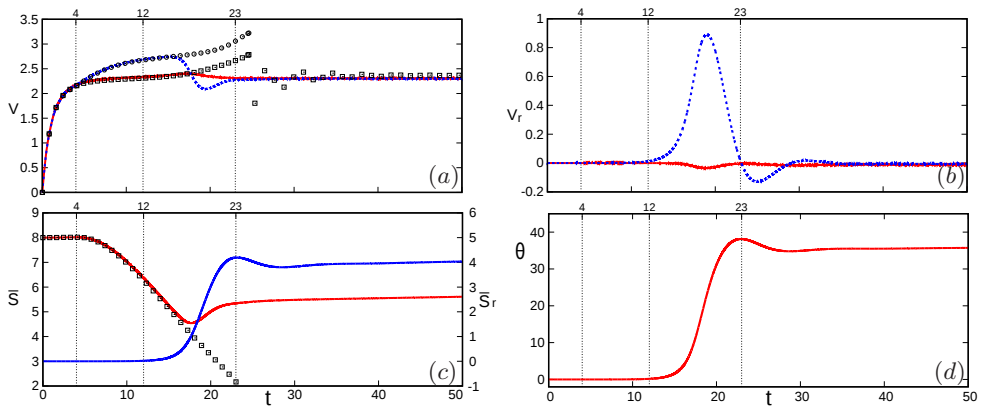}
  \caption{Evolution of several characteristics of a bubble pair with $(Ga, Bo)=(30, 0.3)$ undergoing an ASE interaction. $(a)$: vertical velocity component of the LB (solid red line) and TB (dashed blue line); $(b)$ same for the horizontal component; $(c)$: vertical (red line, left axis) and horizontal (blue line, right axis) separations; $(d)$ inclination of the line of centres with respect to the vertical. Open squares and circles in $(a)$ and $(c)$: axisymmetric prediction.} 
 \label{f4.4.2}
 \vspace{-2mm}
\end{figure}

Computations carried out with somewhat larger values of the Galilei number ($15\leq Ga\leq30$) but still fairly low Bond numbers reveal a drastically different evolution of the bubble pair. In what follows, we select the case $(Ga,\,Bo)=(30,\,0.3)$, \textit{i.e.} $Mo=3.3\times10^{-8}$, as typical of this regime to discuss the corresponding dynamics. With these parameters, an isolated bubble follows a rectilinear path \citep{cano2016paths,Cano2016} and its characteristic rise Reynolds number and aspect ratio at steady state are $Re=66$ and $\chi=1.62$, respectively. The corresponding paths, together with the streamlines past the two bubbles, are shown in figure \ref{f4.4.1}. No standing eddy is observed at the back of the bubbles, in line with the flow structure past an isolated bubble with the same parameters \citep{blanco1995structure}. In panel $(b)$, only streamlines emanating from the right half-plane ahead of the LB are shown for $10\leq t\leq16$ in order to help identify the onset of the non-axisymmetric motion. While all streamlines get around the TB within the right half-plane at $t=10$, one of them deviates to the left half-plane at $t=11$, indicating that axial symmetry has just broken. At this stage, the separation between the two bubbles is still large ($\overline{S}\approx6.8$), in contrast with the situation observed in the DKT scenario. According to panel $(a)$, the TB also departs from its original vertical path at $t\approx11$, but the LB is left virtually unaffected by this departure until $t\approx15$. At this point, the TB still stands in the wake of the LB but its equatorial plane has tilted clockwise, which distorts the flow in the gap and makes it significantly asymmetric at the back of the LB. This asymmetry induces a slight anticlockwise tilt of the LB path until $t\approx20$, whereas the TB goes on drifting laterally and escapes completely from the wake of the LB. Then the TB rotates anticlockwise in such a way that its equatorial plane becomes again horizontal, and its drift stops at $t\approx23$, when the lateral separation between the two bubbles is approximately $\overline{S}_r\approx4$. Due to this temporary anticlockwise rotation, the TB slightly drifts back until $t\approx28$. The motion of the bubble pair has remained planar up to this point. However, at $t\approx28$, they both start to drift slightly out of their previous plane of rise. Then, each bubble goes on rising along a straight but slightly inclined path. Both paths being tilted in the same direction but the angle being larger in the case of the LB, the lateral separation weakly but consistently increases over time. As the TB spent part of its potential energy to move out from the wake of the LB during the time period $11\lesssim t\lesssim23$, it never catches up, and the LB remains significantly ahead of the TB in the final configuration.\vspace{0mm} \\ 
\indent Figure \ref{f4.4.2} describes the evolution of the same four characteristics as in figure \ref{f4.3.2} for the above bubble pair. According to the axisymmetric prediction reported in panel $(c)$, the vertical separation decreases monotonically and the two bubbles enter the initial stage of coalescence at $t\approx23$. However, the three-dimensional evolution reveals a totally different evolution beyond $t\approx11$, although the rise speed of both bubbles and their vertical separation do not exhibit discernible differences with the axisymmetric results up to $t\approx16$. At this time, $V_{TB}$ starts to drop sharply and equals $V_{LB}$ at $t\approx18$. In the meantime, the horizontal velocity of the TB has grown tremendously. Its maximum value, about $40\%$ of $V_{TB}$, is reached at $t\approx19$. The horizontal velocity of the LB reaches its maximum at the same time. However the ratio of the two maxima is less than $4\%$, which confirms that the LB is only marginally disturbed by the lateral drift of the TB. This situation dramatically differs from that depicted in figure \ref{f4.3.2}$(b)$, where the two $V_r$ maxima have almost the same magnitude. Moreover, in the present ASE scenario, the maximum of $V_r$ for the TB is typically three times larger than the maxima encountered during the DKT-type interaction. As figure \ref{f4.4.2}$(c)$ indicates, the horizontal separation has already grown up to $\overline{S}_r\approx2$ by the time $V_r$ reaches its maximum (\textit{i.e.} the TB has left the wake of the LB), and doubles during the next four time units at the end of which the horizontal velocity of the TB vanishes ($t\approx23$). From $t\approx18$ to $t\approx23$, $V_{TB}$ has dropped below $V_{LB}$, which makes the vertical separation re-increase, from $\overline{S}\approx4.5$ at $t=17.5$ to $\overline{S}\approx5.3$ at $t=23$. The $25\%$ drop of $V_{TB}$ from  $t\approx16$ to $t\approx19.5$ emphasizes the fact that a substantial fraction of the kinetic energy of the liquid displaced by the TB has been spent in the meantime to move it laterally and balance the rate of work of the corresponding sideways force. 
 In the next stage ($23\leq t\leq28$), the horizontal velocity of the TB changes sign and reaches a minimum of about $-0.06V_{TB}$. This negative $V_r$ results in the reversed lateral drift already discussed in connection with figure \ref{f4.4.1}. The horizontal  velocity still describes some damped oscillations before the vertical and horizontal separations stabilize and reach values close to $5.6$ and $4$, respectively. Meanwhile, the inclination angle of the bubble pair stabilizes at $\theta\approx36^\circ$. As already pointed out, this configuration is not entirely steady as the slight nonzero slopes of the curves in the right part of figure \ref{f4.4.2}$(c)$ confirm. In other words, the vertical and transverse components of the interaction force driving the relative position of the two bubbles have not completely vanished yet. However, the remaining values of these components are very small, so that it takes an extremely long time for the system to reach a true steady state.
This slow final evolution may be connected to the findings of \cite{hallez2011interaction} for spherical bubbles. Their figure 9$(c)$ indicates that, provided $Re>25$, the two bubbles repel each other whatever their separation distance when the inclination of their line of centres is less than a critical angle $\theta_{c}\approx53^\circ$. However, for $30^\circ<\theta<\theta_c$, this repulsive separation-dependent effect is very weak for $\overline{S}\gtrsim5.0$, which corresponds to the present situation. That the equilibrium angle is close to $37^\circ$ for $Ga=30,Bo=0.3$ instead of the above value for spherical bubbles is likely an effect of the significant oblateness of the bubbles considered here. 
\subsection{Head-on collision and coalescence}\label{sec4.5}
\begin{figure}
\vspace{5mm}
  \centerline{\includegraphics[width=0.9\textwidth]{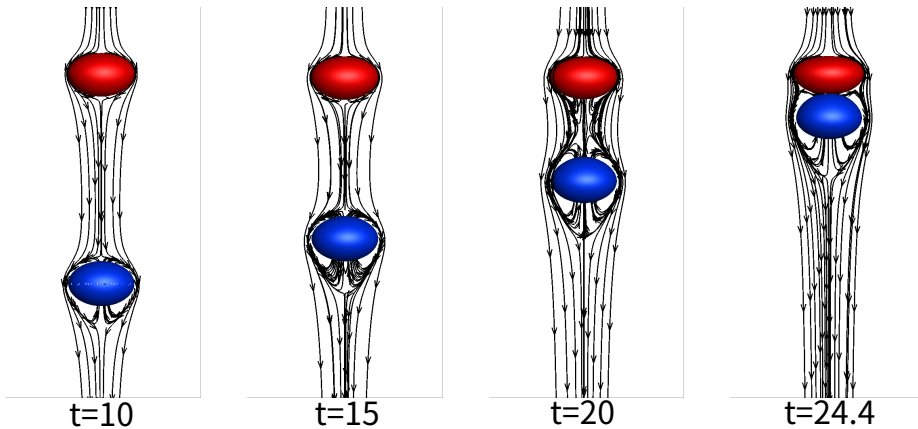}}
  \caption{\color{black} Pre-coalescence dynamics of a bubble pair with $Ga=20, Bo=0.5$; the streamlines in the cross-sectional plane are defined in the reference frame of the leading bubble. }
\label{f4.5.0}
\vspace{0mm}
\end{figure}
\begin{figure}
   \vspace{5mm}
   \centerline{\includegraphics[width=0.9\textwidth]{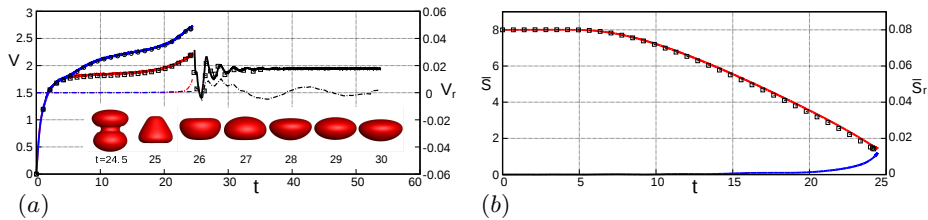}}
    \vspace{-49.5mm}
    \vspace{53mm}
  \caption{\color{black} Evolution of some characteristics of a bubble pair with $Ga=20, Bo=0.5$ undergoing coalescence. $(a)$: vertical (solid lines, left axis) and horizontal (dash-dotted lines, right axis) components of the bubble velocity, with the red, blue and black lines corresponding to the LB, TB and final bubble, respectively; $(b)$: vertical (red line, left axis) and horizontal (blue line, right axis) components of the separation. Open squares and circles: axisymmetric prediction. In $(a)$, the shape of the resulting bubble is shown at several successive time instants during the transient $24.5\leq t\leq30$ following coalescence.}
        \vspace{-31mm}
  \hspace{8mm}$(a)$\hspace{57mm}$(b)$
    \vspace{23mm}\\
\label{f4.5.1}
\end{figure}

As figures \ref{f4.2.1} and \ref{f4.2.2} revealed, increasing the Bond number beyond the $Ga$-dependent threshold $Bo_c(Ga)$ leads to the numerical coalescence of the two bubbles. Most of the time, this coalescence is initiated by a head-on approach, but in some cases the bubbles may also approach each other in an asymmetric manner; \textit{e.g.} the case $(Ga,\,Bo)=(30,\,0.5)$ in figure \ref{f4.2.2}, which corresponds to near-threshold conditions. Actually, the approach configuration depends on the time elapsed since the bubble pair was released. As discussed in \S\,\ref{mechanisms}, the attractive effect of the LB wake increases with the bubble deformation, owing to the direct relation between the interface curvature and the strength of the surface vorticity. This is why, for a given $Ga$, the time at which the two bubbles collide decreases as $Bo$ increases. This leaves less time for non-axisymmetric disturbances to grow before the collision if $Bo$ is significantly larger than  $Bo_c(Ga)$, favouring the head-on configuration. In the case of strongly deformed bubbles, the lift reversal mechanisms discussed in appendix \ref{vortrev} also act to increase the stability of the in-line configuration. 
\color{black} Here we select conditions $(Ga,\,Bo)=(20,\,0.5)$, \textit{i.e.} $Mo=7.8\times10^{-7}$, as an archetype of the head-on coalescence scenario observed in the moderate-Reynolds-number range ($Re\approx35$). As expected from the previous discussion, figure \ref{f4.5.0} shows that 
the two bubbles rise almost in a straight line before they numerically coalesce (a tiny lateral deviation of the LB actually takes place in the late stage of the approach, see figure \ref{f4.5.1}$(b)$). As far as the bubbles evolve independently, their deformation is quite large, characterized by aspect ratios $\chi\approx1.55$ and $\chi\approx1.48$ for the LB and TB, respectively.  In both cases, the aspect ratio is smaller than the critical value $\chi_c=1.65$ beyond which a standing eddy develops at the back of an isolated bubble \citep{blanco1995structure}. This is why no such structure is present in figure \ref{f4.5.0}$(a)$. Beyond $t\approx15$, the shape of the two bubbles changes significantly. On the one hand, the front part of the LB flattens, due to the proximity of the TB which makes its rise speed increase (see figure \ref{f4.5.1}$(a)$). On the other hand, the front part of the TB becomes more rounded, owing to the suction induced by the wake of the LB. Finally, the TB catches up with the LB, leaving only a thin liquid film in the gap and making the two bubbles behave essentially as a bluff compound body.\\
\indent 
After coalescence takes place (see below), figure \ref{f4.5.1}$(a)$ indicates that the resulting bubble first undergoes a series of large-amplitude oscillations until it relaxes to an oblate shape with an aspect ratio close to $2.0$ and a significant fore-aft asymmetry. Due to volume conservation, the radius of this bubble is $2^{1/3}\approx1.26$ times that of the initial bubbles, so that its characteristic parameters are $Ga=2^{1/2}\times20\approx28.3$ and $Bo=2^{2/3}\times0.5\approx0.8$, respectively. Under such conditions, the computations of \cite{cano2016paths} predict that an isolated bubble still rises vertically, although it is close to the transition to a non-straight path. Present observations are in line with these earlier conclusions, since figure \ref{f4.5.1}$(a)$ indicates that the small horizontal velocity component present at the time of coalescence subsequently decreases over time. \color{black}  
\\
\begin{figure}
\vspace{5mm}
  \centerline
  {\includegraphics[width=0.9\textwidth]{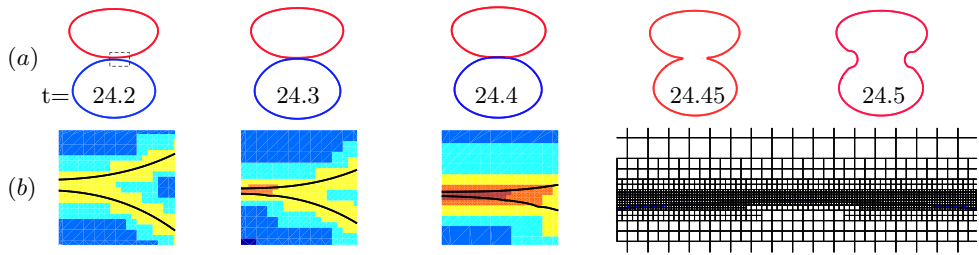}}
\vspace{1mm}
  \caption{Numerical coalescence process. \color{black}
  $(a)$: cross-sectional bubble shapes; $(b)$: zoom on the dashed rectangle in the left panel of $(a)$, showing the successive grid refinements and the grid distribution in the near-contact region just before coalescence ($t=24.4$). In the first three panels of $(b)$, the thin black line indicates the interface and the colour scale spans six grid levels, from $\Delta=R/8.5$ (dark blue) to $\Delta=R/272$ (brown). }
  \vspace{-48.5mm}
  \hspace{0mm}$(a)$\\
  
 \vspace{-2mm}\hspace{4.5mm}t=\hspace{3mm}$24.2$\hspace{18.5mm}$24.3$\hspace{20.5mm}$24.4$\hspace{20mm}$24.45$\hspace{18mm}$24.5$\\
 
\vspace{5mm}\hspace{0mm}$(b)$
\label{f4.5.3}
\vspace{28mm}
\end{figure} \indent 
Returning to the near-coalescence stage, figure \ref{f4.5.3} shows how the interface topology evolves in the stages that just precede and follow numerical coalescence. \color{black} In its late stage, the film is almost flat. No dimple has formed at its periphery, unlike what is customarily observed with coalescing drops \citep{Hartland1968,Hartland1969,Jones1978}. This is because, for low-to-moderate $Bo$ and fully mobile interfaces, a dimple forms only when the film has thinned down by several orders of magnitude \citep{Yiantsios1989}. Indeed, with the same grid resolution, we observed a clear dimple for Bond numbers $\gtrsim1$. Numerical coalescence takes place at $t\approx24.44$. Considering that the film starts to form when the gap is $0.5R$-thick on the symmetry axis, one can estimate that the computation tracks the drainage process during approximately $1.15$ time units.  The arguments discussed in appendix \ref{coal} may then be used to estimate the time by which the film would actually rupture under real conditions. Figure \ref{f4.5.1}$(a)$ indicates that the dimensionless approach velocity $\overline{V}_a=V_{TB}-V_{LB}$ of the two bubbles is approximately $0.5$ during the early stage of the drainage. From this, the approach Weber number, $We_a=\overline{V}_a^2Bo$, and the approach capillary number, $Ca_a=\overline{V}_aBo/Ga=\mu V_a/\gamma$, are found to be $0.125$ and $0.0125$, respectively. Then, for nearly-spherical bubbles, the estimate (\ref{CH}) predicts a dimensionless inertial drainage time $\overline{T}_{di}\approx0.27$, while (\ref{Visc}) (once properly transposed to the case of two identical bubbles) predicts a viscous drainage time $\overline{T}_{dv}\approx2.5$. Hence, the drainage is controlled by viscous effects and the limited resolution would shorten it by $\approx1.35$ time units, would the initial bubble oblateness be small. Given that the tandem rises with the average velocity $\overline{V}_m=\frac{1}{2}(V_{TB}+V_{LB})\approx2.4$, this implies that the vertical position at which coalescence actually happens would be underestimated by $3.25$ bubble radii in this limit. The non-negligible bubble oblateness certainly increases the actual drainage time. For instance, the actual value of the parameter $k_i$ in (\ref{CH}) is $2.05$ for $\chi=1.5$ \citep{Duineveld1994}, so that the correct estimate for the inertial drainage time is  $\overline{T}_{di}\approx0.5$. The quantitative influence of the bubble oblateness on the viscous drainage time is unknown but presumably increases also significantly $\overline{T}_{dv}$.\\
\indent As the last two snapshots in figure \ref{f4.5.3}$(a)$ show, the neck radius of the resulting bubble grows very fast after coalescence happens. More specifically, the growth law (not shown) is found to be $(t-t_{coal})^{0.36}$ over the very first 
stage following the coalescence time instant $t_{coal}$. This is somewhat slower than the $(t-t_{coal})^{1/2}$ self-similar behaviour observed in the experiments of \cite{Paulsen2014} and confirmed theoretically and numerically by \cite{Munro2015} and \cite{Anthony2017}, respectively. However \cite{Paulsen2014} noticed that the growth rate reduces gradually once the neck radius is larger than $0.3R$. Here, due to the limitations inherent to minimum cell size $\delta_{min}=R/272$ and to the numerical procedure used to let the initial two interfaces merge, we barely observe the neck before it reaches this radius, which is certainly the reason for the above slower growth rate. \color{black}

\section{Influence of deformation and initial conditions}\label{sec5}


\subsection{Influence of bubble deformation in the DKT and ASE regimes}\label{sec5.1}

\begin{figure}
\vspace{5mm}
  \centerline{\includegraphics[width=0.8\textwidth]{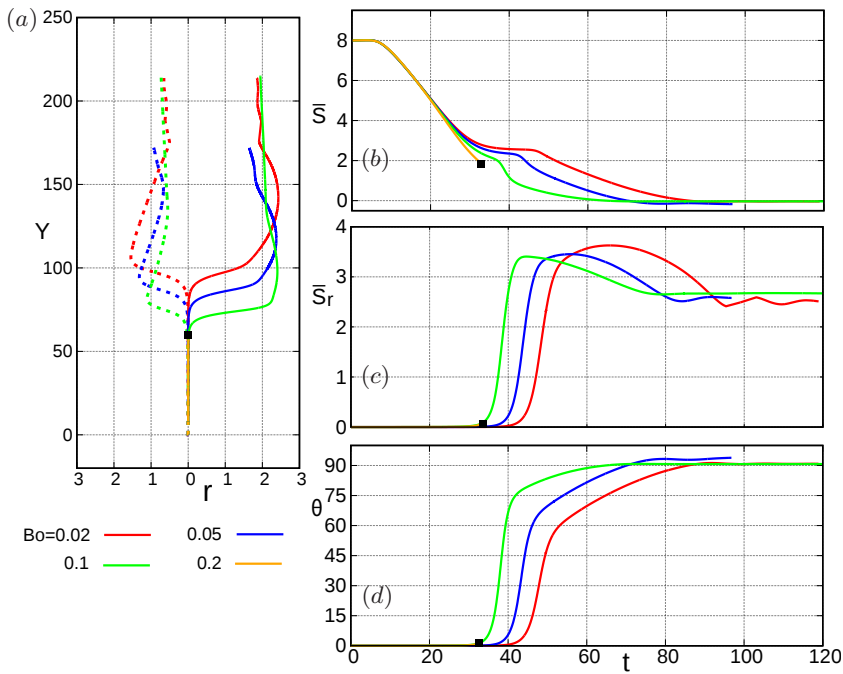}}
  \vspace{1mm}
  \caption{Variations with the Bond number of some path characteristics for $Ga=10$. $(a)$ front view of the path (the solid and dashed lines correspond to the LB and TB, respectively); $(b)$ vertical separation; $(c)$ horizontal separation; $(d)$ inclination of the line of centres. The square symbol denotes the position/time at which coalescence takes place for $Bo=0.2$.}
     \vspace{-87mm}
     
  \hspace{58mm}$(b)$\\

   \vspace{21.5mm} 
   
    \hspace{58mm}$(c)$\\
        \vspace{22mm} 
        
     \hspace{57mm} $(d)$\\
 \vspace{-83mm} 
 
    \hspace{10mm} $(a)$\\
     \vspace{98mm} 
\label{f5.1.0}
\end{figure}

\begin{figure}
\vspace{5mm}
  \centerline{\includegraphics[width=0.8\textwidth]{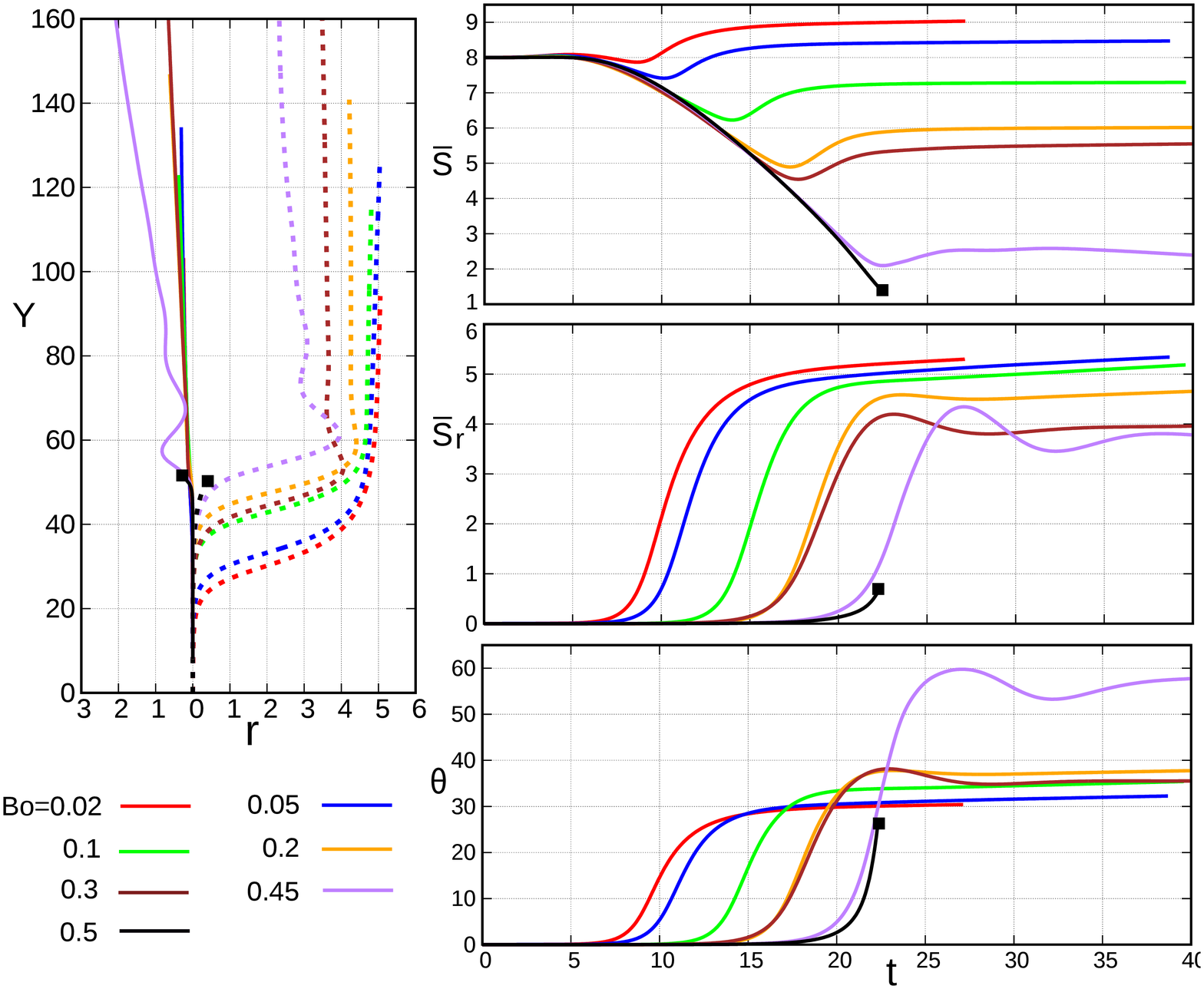}}
  \vspace{1mm}
  \caption{Variations with $Bo$ of some path characteristics for $Ga=30$. $(a)$ front view of the path (the solid and dashed lines correspond to the LB and TB, respectively); $(b)$ vertical separation; $(c)$ horizontal separation; $(d)$ inclination of the line of centres. In $(a)$, the square symbol indicates the position/time at which coalescence takes place for $Bo=0.5$.}
     \vspace{-88mm}
     
  \hspace{58mm}$(b)$\\

   \vspace{22mm} 
   
    \hspace{58mm}$(c)$\\
        \vspace{22mm} 
        
     \hspace{57mm} $(d)$\\
 \vspace{-84mm} 
 
    \hspace{10mm} $(a)$\\
     \vspace{95mm} 
\label{f5.1.1}

\end{figure}

\begin{figure}
\vspace{5mm}
  \centerline{\includegraphics[width=0.9\textwidth]{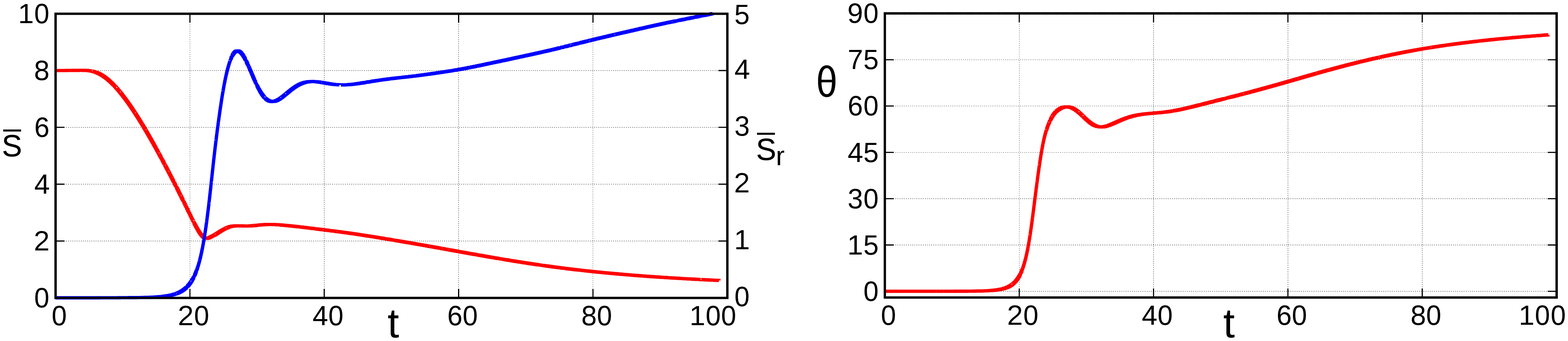}}
  \vspace{-8mm}
  \hspace{12mm}$(a)$\hspace{60mm}$(b)$\\
  \vspace{4mm}

  \caption{Evolution of some characteristics of a bubble pair with $(Ga, Bo)=(30,0.45)$. $(a)$: vertical (red line) and horizontal (blue line) separation distances; $(b)$ inclination of the line of centres.}
\label{f5.1.2}
\vspace{0mm}
\end{figure}

Figure \ref{f5.1.0} shows how several characteristics of the path of bubble pairs with $Ga=10$ vary with the Bond number in the range $0.02\leq Bo\leq0.2$, \textit{i.e.} $8\times10^{-10}\leq Mo\leq8\times10^{-7}$. The DKT scenario is observed for the lower three values of $Bo$, while the pair with $Bo=0.2$ eventually experiences head-on coalescence. \color{black} For lower $Bo$, the interaction always leads to the side-by-side configuration \color{black} and the lateral separation stabilizes at a value close to $2.5$. \color{black} While the Bond number does not have any noticeable influence on the final path in this regime, it affects the critical time by which the tumbling process sets in. \color{black} Owing to the connection between the bubble oblateness and the amount of vorticity produced at its surface, the more oblate the LB is the stronger the attractive wake effect is. Therefore, at a given time, the larger $Bo$ the shorter the separation between the two bubbles. From figures \ref{f5.1.0}$(b)-(d)$, it may be inferred that  the axial symmetry of the flow breaks down when $\overline{S}=\overline{S}_c\approx2.6$, a critical value reached in a shorter time as $Bo$ increases.\\
\indent Figure \ref{f5.1.1} shows the same path characteristics for $Ga=30$ and Bond numbers ranging from $0.02$ to $0.5$, \textit{i.e.} $9.9\times10^{-12}\leq Mo\leq1.5\times10^{-7}$. Here, all cases with $Bo\leq0.45$ correspond to the ASE scenario, whereas the two bubbles coalesce in an asymmetric way for $Bo=0.5$ (see figure \ref{f4.2.2}). In all non-coalescing cases, it is found that the larger $Bo$ the smaller the long-term lateral deviation of the TB (figure \ref{f5.1.1}$(a)$). Hence, the long-term lateral separation decreases with $Bo$ (figure \ref{f5.1.1}$(b)$), since the LB hardly deviates from its initial vertical path except for $Bo=0.45$. 
The vertical separation $\overline{S}$ is also seen to decrease with the Bond number. This is merely a geometrical consequence of the decrease of $\overline{S}_r$. Indeed, the shorter the time spent by the TB in its lateral motion, the shorter the slowing down of its vertical motion, hence the smaller the increase of the vertical separation during the lateral escape stage.\\
\indent  The long-term inclination of the bubble pair exhibits more complex variations. First, it increases with the Bond number up to $Bo=0.2$. Then it slightly decreases for $Bo=0.3$ until it experiences a sharp increase for $Bo=0.45$. In the whole range $Bo\leq0.3$, the long-term inclination angle stands in the range $30^\circ<\theta<40^\circ$, in agreement with the experimental findings of \cite{Kusuno2015} obtained under similar conditions ($Re<150$). The situation corresponding to $Bo=0.45$ is specific. Indeed, the two bubbles are close to coalescing asymmetrically at some point ($\overline{S}\approx2$ at $t=22.5$). This is why the lateral motion of the TB significantly disturbs the subsequent motion of the LB which is seen to experience a series of damped oscillations before rising along a rectilinear, slightly inclined path. Moreover, since the vertical separation of the two bubbles is small just before the TB starts to drift laterally ($\overline{S}\approx3.8$ at $t=18$), the inclination of tandem after this drift has been completed is significantly larger ($\theta\approx60^\circ$) than those found with smaller $Bo$. We shall come back to the long-term evolution of this bubble pair later.  \\
\indent Beyond these geometric features, the main information provided by figure \ref{f5.1.1} is that the smaller $Bo$ is the earlier the side escape of the TB starts. This is in stark contrast with the observations made on figure \ref{f5.1.0} for the DKT regime, where bubble deformation is found to promote the tumbling process. Moreover, considering the critical time at which $\overline{S}_r$ and $\theta$ depart from zero for the various $Bo$ reveals that the corresponding vertical separation does not keep a constant value (figures \ref{f5.1.1}$(b)-(d)$). Rather it decreases from $\overline{S}_c\approx8$ for $Bo=0.02$ to $\overline{S}_c\approx5.8$ for $Bo=0.3$. These values are significantly larger than the equilibrium separation predicted in the axisymmetric configuration which, according to figure \ref{f4.1.2}, ranges from $\overline{S}_e\approx6.1$ for $Bo=0.02$ to $\overline{S}_e\approx4.3$ for $Bo=0.3$. Moreover, in all cases, this critical separation is much larger than that at which tumbling is found to start in the DKT regime. This finding indicates that, unlike the breakdown of the axial symmetry in the DKT scenario, the ASE mechanism is driven by a long-range interaction. This is in line with the conclusion of \cite{Yin2008} who simulated the motion of buoyancy-driven suspensions of spherical non-coalescing bubbles at $Re\approx10$.\\
\indent To get additional insight into the long-term behaviour of the bubble pair, figure \ref{f5.1.2} shows the evolution of the separation and inclination angle over longer times for the two sets $(Ga,\,Bo)=(30,\,0.3)$ and $(Ga,\,Bo)=(30,\,0.45)$ , \textit{i.e.} $Mo=3.3\times10^{-8}$ and $Mo=1.1\times10^{-7}$, respectively. In the former case, the two components of the separation are seen to slightly increase until the end of the computation but the inclination angle changes by less than $1^\circ$ over the last forty time units, reaching the `asymptotic' value $\theta\approx36^\circ$. Conversely, in the latter case, the two components of the separation exhibit significant variations throughout the computation. The vertical separation gradually tends to zero while $\overline{S}_r$ goes on increasing even for $t\approx100$. At this final time, $\theta$ is close to $83^\circ$ and is still gently increasing, so that there is little doubt that the system eventually reaches a perfect side-by-side configuration. To understand why the two bubble pairs behave so differently on the long term, it must first be noticed that, at the end of the lateral drift of the TB ($t\approx40$), the distance $\mathscr{S}=(\overline{S}^2+\overline{S}_r^2)^{1/2}$ between the two centroids is approximately $6.8$ for $Bo=0.3$ and $4.5$ for $Bo=0.45$, the Reynolds numbers being approximately $66$ and $60$, respectively. For spherical bubbles, the results of \cite{hallez2011interaction} indicate that, for $Re=100$, the torque acting on the tandem is negligibly small for $\theta\gtrsim25^\circ$ when $\mathscr{S}\gtrsim6$ but keeps significant values whatever $\theta$ for $\mathscr{S}=4.5$. Although bubble deformation is large in the two sets considered in figure \ref{f5.1.2}, this is a strong indication that, after the TB has drifted laterally, only the pair with $Bo=0.45$ still experiences a noticeable torque. Consequently only this pair may reach the side-by-side configuration in a reasonable time. According to figure \ref{f5.1.1}, $\mathscr{S}$ decreases with $Bo$. Hence, it may be concluded that in the ASE scenario, the distance between the two centroids after the TB has completed its drift is usually too large for the torque to remain significant. Therefore the inclination of the tandem remains almost unchanged in the subsequent stages. Only bubbles whose deformation is close to the coalescence threshold (here $Bo\approx0.5$) escape this rule, since the vertical separation of the corresponding pairs is very small when the side escape of the TB takes place.\\
 \indent Interestingly, for $Bo=0.45$, $\overline{S}_r$ goes on increasing at the end of the computation, suggesting that the interaction force is still nonzero and repels the two bubbles. At the same $Re$ and $\mathscr{S}$, the computational results of \cite{Legendre2003} indicate that the interaction force between two spherical bubbles rising side-by-side is attractive. The difference between the two situations may be understood by noting once again that wake effects are much stronger for $Bo=0.45$ than for $Bo=0$. Since these effects are repulsive in the side-by-side configuration, the critical Reynolds number at which the interaction force switches from repulsive to attractive is significantly larger in the present case, which explains the observed behaviour.
 \begin{figure}
\vspace{-55mm}
  \centerline{\includegraphics[width=0.9\textwidth]{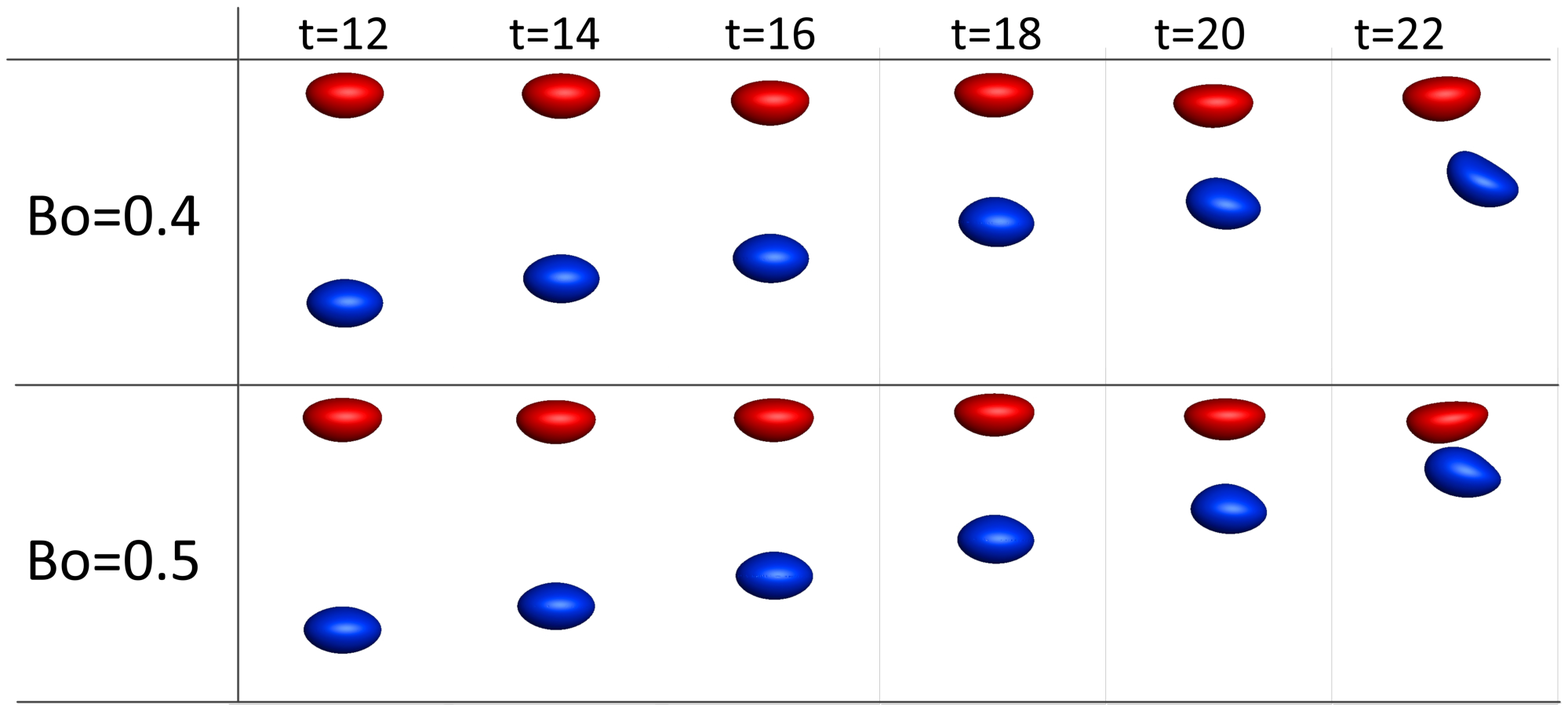}}
      \vspace{-118mm}
  \centerline {\includegraphics[width=0.92\textwidth]{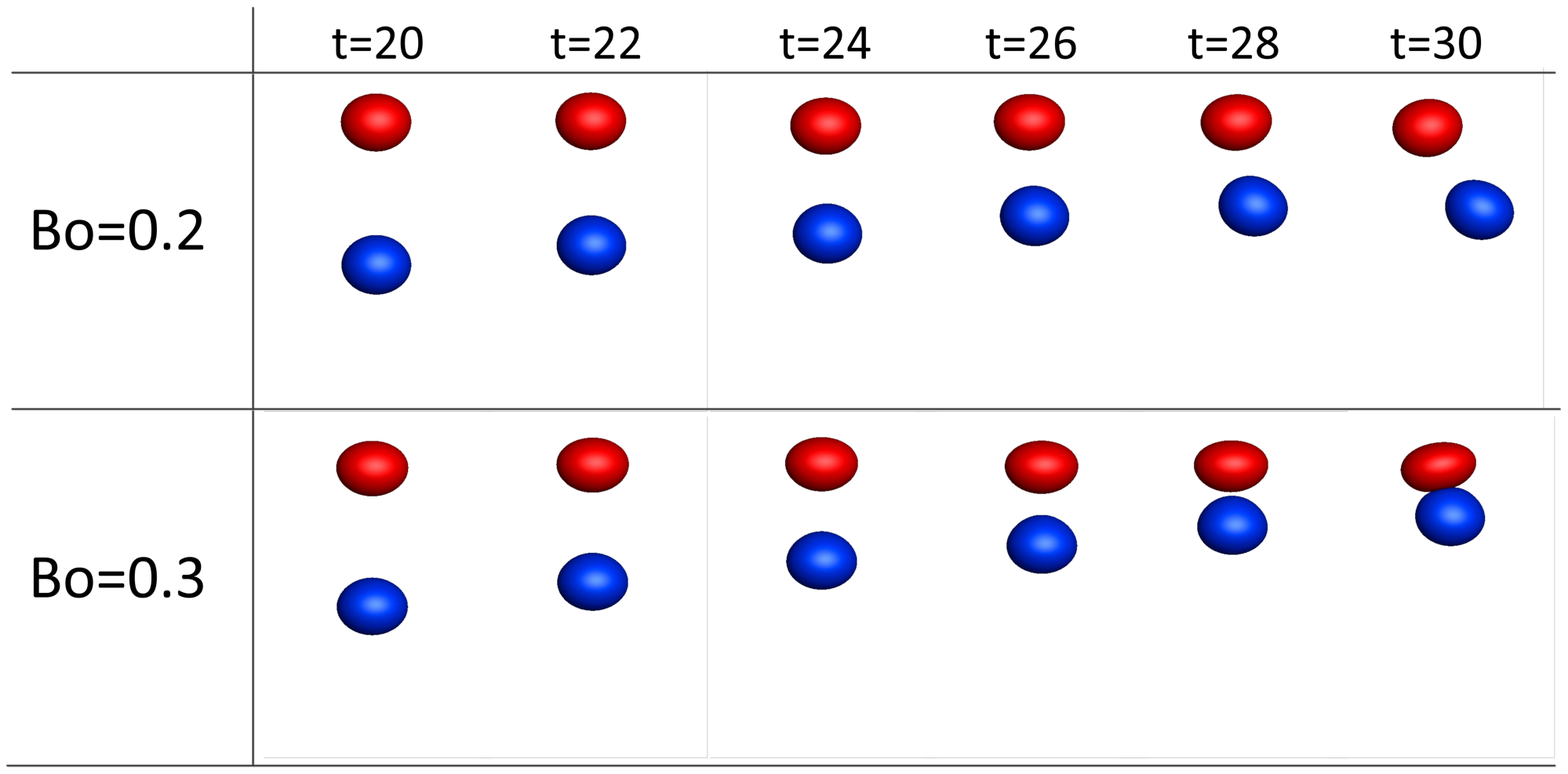}}
  \vspace{-145mm}
    \hspace{0mm}$(a)$
       \vspace{52mm}\\
         \hspace{30mm}$(b)$
    \vspace{27mm}
  \caption{Evolution of the bubble shapes and relative positions for near-critical conditions. $(a)$: $Ga=30$; $(b)$: $Ga=15$.}
\label{shapes}
   \vspace{0mm}
\end{figure}
\subsection{Role of the TB shape in near-critical conditions for coalescence}
\label{shapereverse}
The global geometric indicators reported in figure \ref{f5.1.1} reveal how the two bubbles get closer to coalescence as the Bond number increases. Nevertheless it is of interest to examine also how the bubble shapes vary with $Bo$ close to the corresponding threshold, especially with respect to the potential influence of the bubble distortion on the magnitude and even the direction of the sideways force acting on the TB through the two mechanisms reviewed in appendix \ref{vortrev}. Figure \ref{shapes} displays the evolution of the shapes and relative positions of the two bubbles for near-critical conditions at $Ga=30$ and $15$, respectively. It is striking that, for both values, the TB shape exhibits almost no left/right asymmetry until the late stage of the interaction. This is a strong indication that the $A$-mechanism discussed in appendix \ref{vortrev} cannot reduce significantly the sideways force. Similarly, the flow characteristics are such that the $S$-mechanism cannot take place. For instance, with $Ga=30$ and $Bo=0.4$, the aspect ratio of the TB at $t=14$ is approximately $1.6$ and its Reynolds number is close to $80$. For these parameters, the path of an isolated bubble rising in a fluid at rest is stable and the numerical results of \cite{adoua2009reversal} (their figure 5) indicate that finite-Reynolds-number effects decrease the lift force by only $15\%$ with respect to the inviscid prediction. So, it can be concluded that within the parameter range considered in the present study, the lateral migration of the TB is merely driven by the standard shear-induced mechanism. Hence, close to the threshold, what makes the difference between non-coalescing and coalescing situations is essentially the enhancement of the attractive wake effect as the TB becomes increasingly oblate. For a sufficiently large $Bo$, this attractive effect becomes so strong that it delays the occurrence of the lateral instability of the TB to such an extent that, although this bubble subsequently migrates `normally', its migration lasts for a too short time to prevent it from hitting the LB. 
\subsection{Influence of an initial angular deviation}
\label{devi}
Despite sophisticated bubble release systems, tiny initial lateral deviations can hardly be avoided in laboratory experiments, yielding small-but-nonzero angular deviations of the tandem with respect to the perfect in-line configuration. This calls for the investigation of the influence of an initial nonzero inclination on the subsequent dynamics. For this purpose, we systematically examined the impact of a nonzero  initial inclination $0^\circ<\theta_0\leq2^\circ$ throughout the $(Ga,\,Re)$-range considered in \S\S\,\ref{sec4.1} and \ref{3D}. \color{black} Note that in all configurations considered in this subsection, the initial horizontal separation between the bubble centres  is several times larger than the finest grid cells located on both sides of the bubbles surface ($4$ times larger for the smallest inclination). This initial configuration results in a `macroscopic' asymmetry of the discretized solution, in which the tiny numerical asymmetries of the pressure field returned by the Poisson solver (see \S\,6.2 and appendix \ref{anum}) play no role. \color{black}\\
\indent For $Ga=10$, we observed that under such conditions, the DKT scenario no longer takes place for $Bo\leq0.1$. Instead, the system follows a clear ASE evolution.  For instance, selecting $(Ga,\,Bo)=(10,\,0.1)$, the TB starts drifting laterally when $\overline{S}\approx4$ (resp. $\overline{S}\approx5$) if $\theta_0$ is set to $1^\circ$ (resp. $2^\circ$), leaving the path of the LB almost unaffected in both cases. Instead of ending up in the side-by-side configuration as it does when $\theta_0=0^\circ$, the tandem then reaches an approximate final inclination of $50^\circ$ (resp. $40^\circ$). That $\theta_0$ has such a dramatic influence on the existence of the DKT regime and the late geometry of the tandem is at variance with observations reported for rigid bodies falling in tandem. Indeed, experiments performed with short cylinders \citep{brosse2014interaction} 
indicate that the DKT regime is still observed when a small initial lateral offset is imposed to the trailing body. The key difference with a bubble pair is the relative magnitude of attractive effects which is much stronger for rigid bodies, owing to the different vorticity generation mode at the body surface. Moreover, for a given body shape, shear-induced lift effects are significantly weaker for rigid bodies with Reynolds numbers of $\mathcal{O}(10)$ or larger, owing to the presence of large separated regions \citep{kurose1999drag}. These two factors make the DKT configuration much more sensitive to small lateral deviations in the case of nearly-spherical bubbles.\\
\begin{figure}
\vspace{5mm}
  \centerline{\includegraphics[
  width=0.96\textwidth]{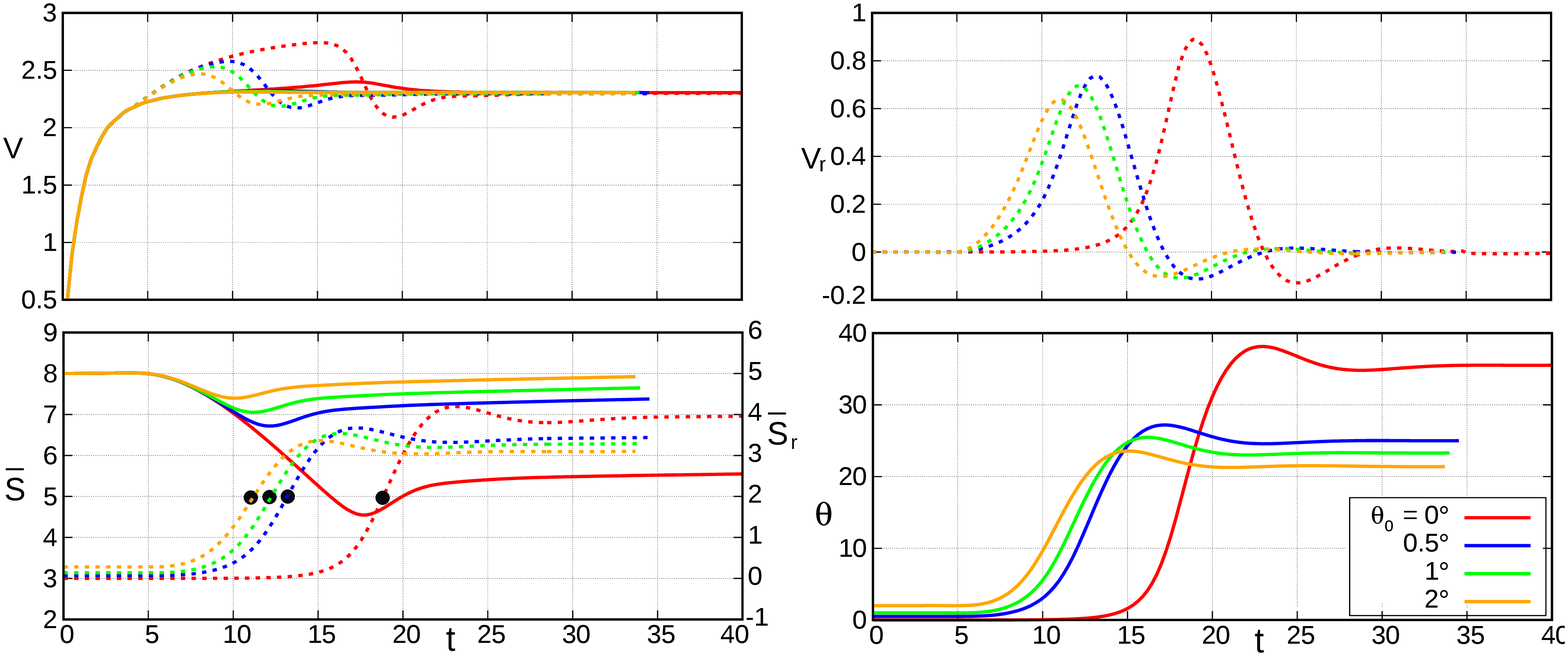}}
  \vspace{-34.2mm}
  \hspace{9mm}$(a)$\hspace{63mm}$(b)$\\
  \vspace{28mm}
  \caption{Evolution of several characteristics of a bubble pair with $(Ga, Bo)=(30, 0.3)$ undergoing initial angular deviations $\theta_0$ from $0^\circ$ to $2^\circ$. $(a)$: vertical velocity of the LB (solid line) and TB (dashed line); $(b)$ horizontal velocity of the TB; $(c)$: vertical (solid line, left axis) and horizontal (dashed line, right axis) separations; $(d)$ inclination of the line of centres. In $(b)$, the bullets on the $\overline{S}_r$-curve indicate the location where the maxima of $V_r$ are reached.}
\label{f5.2.4}
  \vspace{-32.5mm}
    \hspace{9mm}$(c)$\hspace{63mm}$(d)$\\
    \vspace{25mm}
\end{figure}
\indent Figure \ref{f5.2.4} provides the evolution of the tandem geometry in the case $(Ga,\,Bo)=(30,\,0.3)$ for initial inclinations $\theta_0$ ranging from $0^\circ$ to $2^\circ$. The system is found to follow the ASE scenario in all cases. However its final geometry strongly depends on $\theta_0$. In particular, as figure \ref{f5.2.4}($d$) shows, the final inclination of the line of centres decreases from $36^\circ$ for $\theta_0=0^\circ$ to $22^\circ$ for $\theta_0=2^\circ$. This is because when $\theta_0$ is nonzero, the TB starts drifting laterally soon after it is released from rest, which is not the case for $\theta_0=0^\circ$. Indeed, with $\theta_0\neq0^\circ$, the initial flow configuration is no longer axisymmetric and the TB faces an asymmetric wake as soon as vorticity generated at the LB surface has been advected downstream over an $O(\overline{S})$-distance. Consequently, the TB experiences a nonzero sideways force much earlier than in the perfect in-line configuration, where this force occurs only after the system has become unstable. A nonzero $\theta_0$ shortens the initial time period after which $\overline{S}_r$ starts to grow, as is made clear in figure \ref{f5.2.4}$(c)$. A direct consequence of this shortening is the fact that the vertical separation at which the lateral drift starts is larger for $\theta_0\neq0^\circ$ ($\overline{S}\approx8$ for $\theta_0=2^\circ$ instead of $\overline{S}\approx6.8$ for $\theta_0=0^\circ$). For this reason, weaker velocity gradients  across the LB wake subsist for $\theta_0\neq0^\circ$ at the position of the TB when it starts drifting, resulting in a weaker sideways force. This makes the maximum of $V_r$ smaller (figure \ref{f5.2.4}$(b)$), yielding a shorter final lateral position, hence a smaller final inclination of the tandem. \\
\begin{figure}
   \vspace{5mm} 
  \centerline{\includegraphics[width=0.8\textwidth]{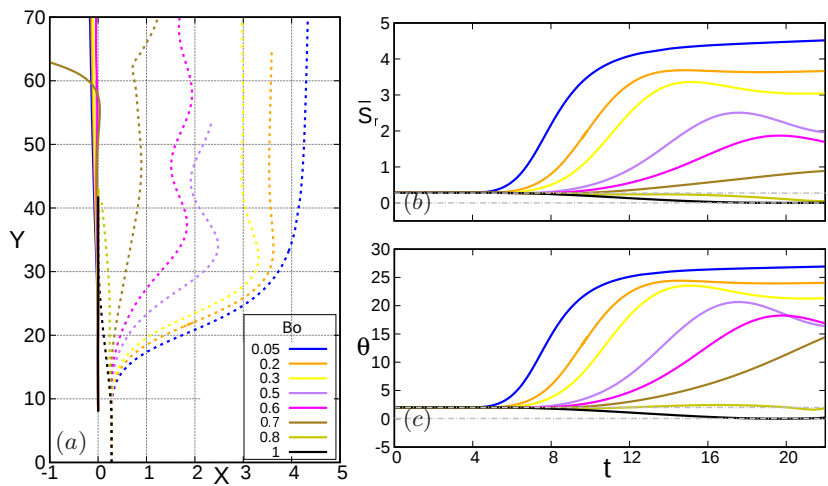}}
  \caption{Variations with $Bo$ of some path characteristics for $Ga=30$ in the presence of an initial angular deviation $\theta_0=2^\circ$. $(a)$ front view of the path (the solid and dashed lines correspond to the LB and TB, respectively); $(b)$ horizontal separation; $(c)$ inclination of the line of centres.}
      \vspace{-55.5mm}
  \hspace{65.5mm}$(b)$\\
  
   \vspace{21.5mm} 
    \hspace{65.5mm}$(c)$\\
    
        \vspace{-4mm}         
      \hspace{18.5mm} $(a)$\\
     \vspace{17mm} 
     \label{f5.2.5}
\end{figure}
\indent Figure \ref{f5.2.5} displays the influence of the Bond number on the evolution of bubble pairs with $Ga=30$ when $\theta_0$ is set to $2^\circ$. This figure is the counterpart of figure \ref{f5.1.1} discussed above for $\theta_0=0^\circ$. All pairs with $Bo<0.7$, \textit{i.e.} $Mo<4.2\times10^{-7}$, are seen to follow a clear ASE scenario. For the aforementioned reason, the final inclinations reported in figure \ref{f5.2.5} are significantly smaller than those observed in figure \ref{f5.1.1}. In addition, they exhibit a marked and consistent decrease with the Bond number, a trend which is absent from figure \ref{f5.1.1}.  For pairs with $Bo\geq0.8$, $\theta$ is found to decrease over time until the two bubbles perfectly align vertically, which eventually forces them to coalesce. The near-critical case $Bo=0.7$  is quite specific, and in many instances similar to the situation encountered for $Bo=0.45$ in figure \ref{f5.1.1}. In this case, the two bubbles are very close to coalescing at $Y\approx58$ in figure \ref{f5.2.6}$(a)$. The corresponding gap is so thin that the flow at the back of the LB is strongly disturbed, forcing the latter to deviate abruptly from its vertical path. This allows the lateral separation to increase beyond the critical value $\overline{S}_{rc}\gtrsim2$ within a short time lapse (not visible in figure \ref{f5.2.6}$(b)$ which is limited to shorter times), allowing the tandem to avoid coalescence. Similar to the evolution displayed in figure \ref{f5.1.2}$(b)$, but through a sharper transition, the inclination of the tandem grows until the side-by-side configuration is reached. Given the small gap at which the transition takes place and the quite symmetric final lateral positions of the two bubbles,  this specific evolution is closer to the DKT scenario than to a standard ASE evolution.\\
\indent The critical Bond number beyond which the two bubbles coalesce stands in the range $0.45-0.5$ in figure \ref{f5.1.1} ($\theta_0=0^\circ$), but is slightly larger than $0.7$ for $\theta_0=2^\circ$. In line with the discussion on figure \ref{f5.2.4}, the reason for this marked increase stems directly from the longer vertical distance over which the shear-induced lift force acts on the TB when $\theta_0\neq0^\circ$, and the slightly shorter lateral distance this bubble has to drift to avoid hitting the LB. These two factors imply that, compared to the reference case, a weaker positive lift force is sufficient to avoid coalescence when $\theta_0$ is nonzero. The two lift reversal mechanisms discussed in appendix \ref{vortrev} being directly linked to the ability of the TB to deform, increasing the Bond number makes the lift force decrease and eventually change sign, other things being equal. Hence, in the presence of an initial angular deviation, coalescence can be avoided over a broader range of distortion of the TB, \textit{i.e.} up to a larger Bond number. \\
\begin{figure}
\vspace{5mm}
  \centerline{\includegraphics[width=0.96\textwidth]{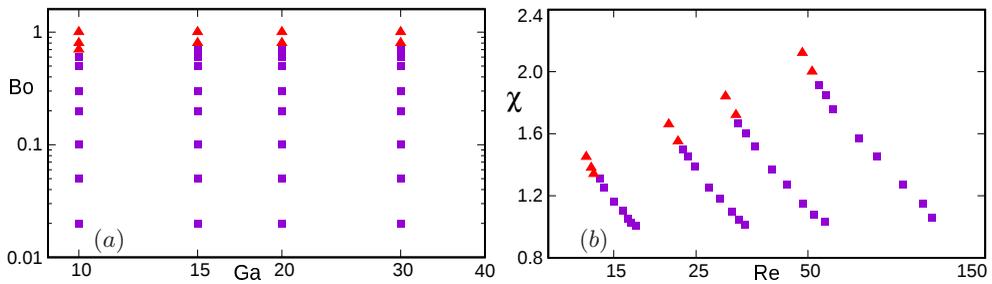}}
  \vspace{-8mm}
  \hspace{14mm}$(a)$\hspace{60mm}$(b)$\\
 \vspace{1mm}
  \caption{Phase diagram showing whether a bubble pair released with an angular deviation $\theta_0=2^\circ$ evolves in the ASE regime (squares) or eventually coalesces (triangles). $(a)$ $(Ga, Bo)$ map; $(b)$ $(Re, \chi)$ representation based on the steady-state values corresponding to the rise of the corresponding isolated bubble. }
\label{f5.2.6}
\vspace{-4mm}
\end{figure}
 \indent The fate of all bubble pairs released with an initial inclination $\theta_0=2^\circ$ over the range $10\leq Ga\leq30,\,0<Bo\leq2.0$ is summarized in figure \ref{f5.2.6}. This figure is the counterpart of figure \ref{f4.2.1} obtained with $\theta_0=0^\circ$. In line with the above discussion, the ASE regime observed when $\theta_0=2^\circ$ exists over a significantly broader range of $Bo$, hence for bubbles with a larger oblateness. For instance, bubbles with aspect ratios $\chi\gtrsim1.3$ (resp. $1.7$) are found to coalesce for $Ga=15$ (resp.  $30$) in the absence of any initial deviation. With $\theta_0=2^\circ$, the corresponding critical aspect ratios raise beyond $1.5$ (resp. $2.0$) and the critical Reynolds numbers are close to $50$ and $120$, respectively.\\
 \begin{figure}
\vspace{-55mm}
  \centerline{\includegraphics[width=0.9\textwidth]{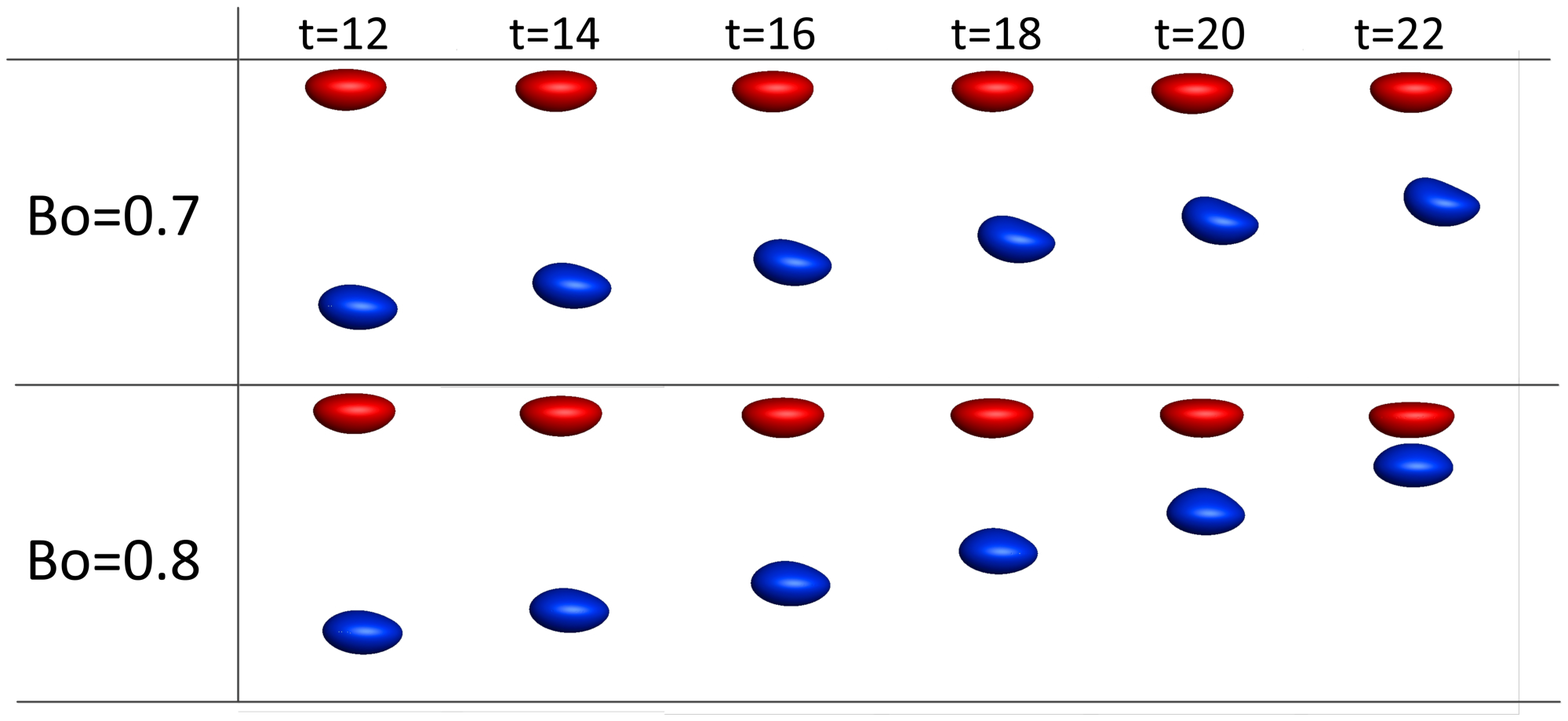}}
         
    \vspace{-61mm}
  \caption{Evolution of the bubble shapes and relative positions at $Ga=30$ under near-critical $Bo$-conditions in the presence of an initial angular deviation $\theta_0=2^\circ$.}
\label{shapes2}
   \vspace{-1mm}
\end{figure}
\indent Figure \ref{shapes2} shows how the bubble shapes and relative positions evolve close to the transition to coalescence for $Ga=30$. Unlike those reported in figure \ref{shapes}, the TB shapes now exhibit a marked left/right asymmetry (although less pronounced, this trend subsists for lower $Ga$). The difference with the strict in-line configuration is that here the TB is fully immersed in an asymmetric flow as soon as the wake of the LB has developed within the gap. Consequently, its shape has to adapt to this asymmetric environment throughout the interaction process. While the TB is seen to migrate toward the right for $Bo=0.7$, a tiny migration toward the left may be identified for $Bo=0.8$, leading unavoidably to coalescence. The discussion in appendix \ref{vortrev} allows the origin of this reverse migration to be readily identified. The aspect ratio of the TB being only $1.85$ for $Bo=0.8$, the $S$-mechanism is not present, since it only takes place beyond a threshold $\chi_{cS}\approx2.2$ ( we observed a reversed migration due to this mechanism by increasing the Bond number beyond unity). 
 In contrast, the egg-like shapes of the TB point to the $A$-mechanism. The capillary number based on the rise velocity of this bubble at $t=14$ only increases from $0.051$ for $Bo=0.7$ to $0.058$ for $Bo=0.8$. However this modest increase turns out to be sufficient for the deformation-induced force directed to the left to exceed the shear-induced lift directed to the right, preventing the lateral escape of the TB.

\subsection{Influence of initial separation}\label{sec5.3}
\begin{figure}
\vspace{5mm}
  \centerline{\includegraphics[width=0.99\textwidth]{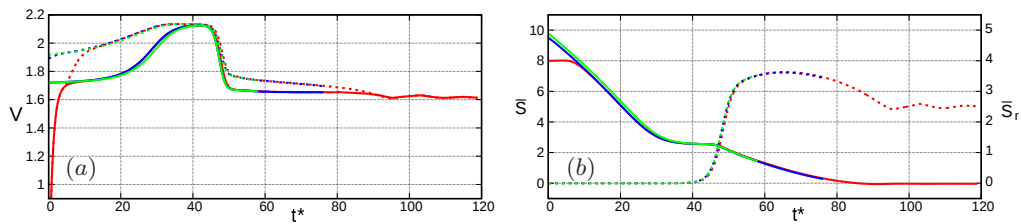}}
  \vspace{-9.7mm}
 \hspace{7mm} $(a)$\hspace{62mm}$(b)$\\
 \vspace{4mm}
  \caption{Evolution of some characteristics of a bubble pair with $(Ga, Bo)=(10, 0.02)$ for different initial separations. $(a)$: rise speed of the LB (solid lines) and TB (dashed lines); $(b)$ vertical (solid lines, left axis) and horizontal (dashed lines, right axis) separations. Red, blue and green lines refer to $\overline{S}_0=8,\,10$ and $12$, respectively. A time shift $t_0=11$ (resp. $t_0=22$) is applied for $\overline{S}_0=10$ (resp. $S_0=12$) and evolutions are plotted \textit{vs.} the modified time $t^*=t-t_0$. 
  }
  \vspace{-4mm}
\label{f5.3.1}
\end{figure}
All simulations considered up to now are based on an initial separation distance $\overline{S}_0=8$. It is relevant to examine how this choice influences the upcoming evolution of the bubble pair. Since the in-line configuration is stable in the head-on coalescence regime, $\overline{S}_0$ is not expected to have an influence in this regime. This is why we consider the influence of $\overline{S}_0$ only in the DKT and ASE regimes. \\
\indent Figure \ref{f5.3.1} displays the evolution of some characteristics of the bubble pair obtained by increasing $\overline{S}_0$ from $8$ to $12$ in the case $Ga=10,\,Bo=0.02$. With $\overline{S}_0=8$, this set of parameters yields the DKT configuration described in figures \ref{f4.3.1} and \ref{f4.3.2}.
Introducing an appropriate $\overline{S}_0$-dependent time shift $t_0$ and setting $t^*=t-t_0(\overline{S}_0)$ allows the $t^*$-evolutions of all quantities for the three initial separations to collapse on a single curve beyond $t^*\approx10$. Hence a DKT scenario yielding the same final configuration is observed whatever the initial separation. This is no real surprise since, for this set of parameters, the in-line configuration becomes unstable only when the two bubbles get very close, which happens only for $t^*\approx40$.

\begin{figure}
\vspace{5mm}
\centerline{\includegraphics[width=0.95\textwidth]{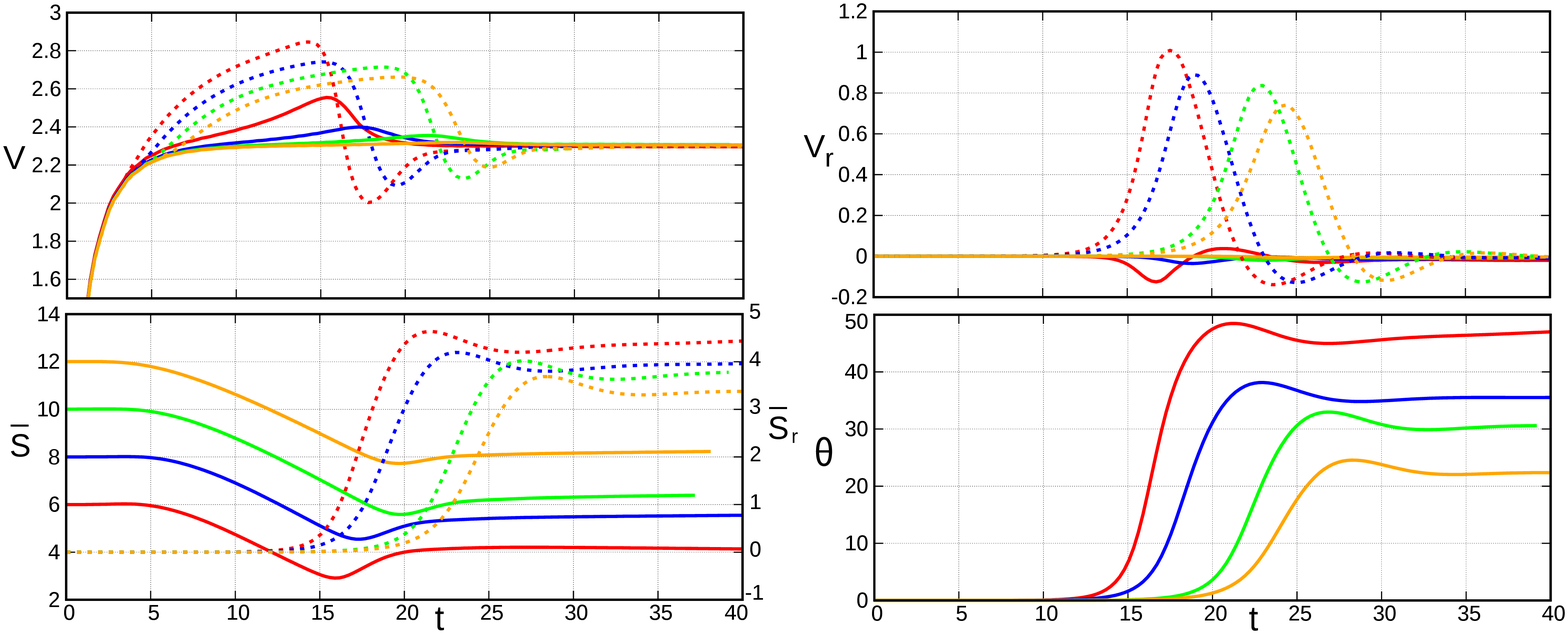}}
 \vspace{-51.5mm}
  \hspace{10mm}$(a)$\hspace{62mm}$(b)$\\
   \vspace{41.5mm}\\
       \caption{Influence of the initial separation $\overline{S}_0$ on several characteristics of a bubble pair with $(Ga, Bo)=(30, 0.3)$. $(a)$: vertical velocity component of the LB (solid line) and TB (dashed line); $(b)$ same for the horizontal component; $(c)$: vertical (solid line) and horizontal (dashed line) separations; $(d)$ inclination of the line of centres. Red, blue, green and orange lines correspond to $\overline{S}_0=6, 8, 10$ and $12$, respectively.} 
     \vspace{-27.5mm}
         \hspace{10mm}$(c)$\hspace{62mm}$(d)$
  \vspace{22.5mm}
\label{f5.3.2}
\end{figure}

\begin{figure}
  \vspace{4mm}
  \centerline{\includegraphics[width=0.98\textwidth]{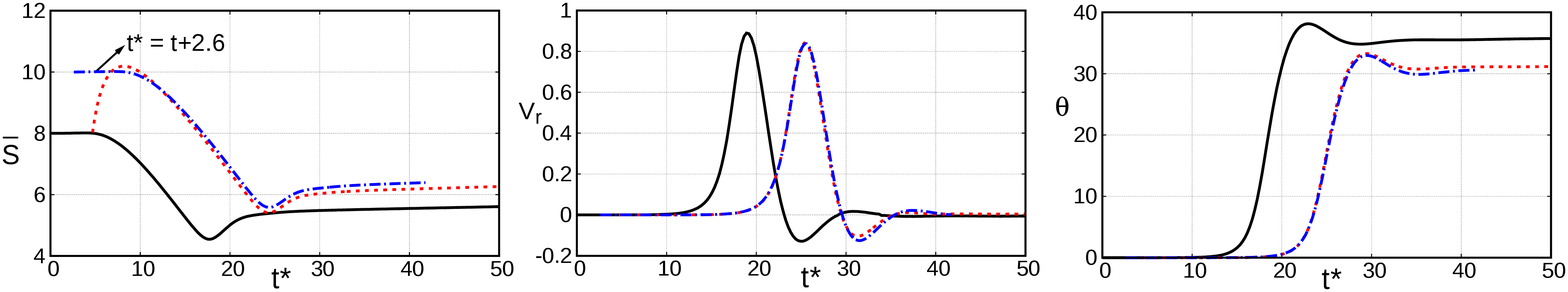}}
  \vspace{-7.5mm}
  \hspace{7mm}$(a)$\hspace{40.5mm}$(b)$\hspace{40.5mm}$(c)$
  \vspace{5mm}
  \caption{Influence of the sequential release of the two bubbles. $(a)$ vertical separation; $(b)$ lateral velocity; $(c)$ inclination angle. Solid line: bubbles are released simultaneously with $\overline{S}_0=8$; blue dash-dotted line: same with $\overline{S}_0=10$ and a time shift $t_0=2.6$; red dashed line: the TB is released from rest after the LB has travelled a distance $\overline{S}(t_0)=8$. 
  }
\label{f5.3.3}
\end{figure}

Figure \ref{f5.3.2} shows how the evolution of the same parameters vary with $\overline{S}_0$ for $Ga=30,\,Bo=0.3$. In this case, the interaction process yields an ASE scenario whatever $\overline{S}_0$. However the final configuration now depends on the initial separation. In particular, the shorter $\overline{S}_0$ the larger the final inclination of the tandem, with $\theta\approx47^\circ$ and $\theta\approx22^\circ$ for $\overline{S}_0=6$ and $\overline{S}_0=12$, respectively. Not unlikely, the larger $\overline{S}_0$ is the longer it takes for the path of the TB to start bending. Examining the evolutions of the vertical separation and those of the rise speed of each bubble reveals that none of these three quantities exhibits a constant value by the time the TB starts drifting laterally. In contrast, it turns out that the difference $V_{TB}-V_{LB}$ is close to $0.35$  whatever $\overline{S}_0$ when this lateral motion sets in. Therefore we conclude that the in-line configuration becomes unstable when the velocity difference between the two bubbles exceeds the above threshold, irrespective of the separation distance at which this threshold is reached.\\
\indent Closely related to the influence of the initial separation is that of the sequential release of the two bubbles in actual laboratory experiments.
Still with $Ga=30,\,Bo=0.3$, we ran a computation in which the TB was allowed to start rising only after the LB reached a dimensionless height $\overline{S}(t_0)=8$ above the point of release. Figure \ref{f5.3.3}$(a)$ displays the corresponding evolution of the vertical separation. Due to the time lapse the TB needs to reach its final rise speed, $\overline{S}$ reaches a maximum close to $10.2$ before the attractive interaction sets in and the separation starts to decrease. This is why we compare the subsequent evolution with that obtained for $\overline{S}_0=10$ when the two bubbles released simultaneously. For this purpose, we introduce an appropriate time shift $t_0$ and define a modified time $t^*=t+t_0$ to make the two evolutions coincide in the early stage of the interaction. 
With this time shift, the two further evolutions of $\overline{S}$ almost superimpose, as do those of the lateral velocity of the TB (figure \ref{f5.3.3}$(b)$) and the inclination angle (figure \ref{f5.3.3}$(c)$). Consequently it can be concluded that the time elapsed in between the release of two successive bubbles merely increases their `initial' separation, defined as the distance that separates them before the wake of the LB starts to influence the rise of the TB. 

\section{Summary and concluding remarks}\label{sec6}

We carried out a series of three-dimensional simulations in order to dissect the physical mechanisms involved in the evolution of a pair of clean, deformable rising bubbles initially released in line. In this first part of the investigation, we focused on the parameter range $10\leq Ga\leq30,\, 0.02\leq Bo\leq2.0$ which corresponds to inertia-dominated regimes in which the path of an isolated bubble remains straight and vertical. We made use of the Basilisk open source code and improved on the original version by implementing a specific AMR improvement \citep{zhang2019vortex} allowing an automatic grid refinement in the gap left between the two bubbles when they come almost in contact. Previous theoretical and computational investigations of this configuration essentially considered spherical bubbles and steady or quasi-steady configurations. By allowing the bubbles to move and deform freely, the present study provides a more realistic description of the hydrodynamic interactions governing the fate of the bubble pair.

To build on a reference case, we first ran axisymmetric time-dependent simulations. Similar to the predictions of \cite{Yuan1994} and \cite{hallez2011interaction} for spherical bubbles, we found that the two bubbles stabilize a finite distance apart, provided their deformation remains moderate. Since the strength of the inertia-induced repulsive interaction force increases with $Ga$, so does the critical Bond number beyond which the bubbles come in contact and eventually coalesce. 
For this reason, the equilibrium distance depends on both $Ga$ and $Bo$, or equivalently, on the Reynolds and Weber numbers. The empirical correlation (\ref{e4.2}) summarizes the corresponding findings. \\
\indent We then turned to fully three-dimensional evolutions. The simulations revealed that, under conditions for which an equilibrium separation distance exists in the axisymmetric case, this configuration is never reached in the three-dimensional case because the in-line arrangement is unstable with respect to non-axisymmetric disturbances. As rationalized long ago \citep{Harper1970}, this is because, provided bubble deformation is moderate, any deviation of the TB from the axis of the LB wake is amplified, owing to the shear-induced lift force that tends to drive it out of the wake.
This lateral drift takes two markedly different forms, depending on the strength of inertial and deformation effects. For $Ga=\mathcal{O}(10)$ and $Bo\lesssim 0.1$, a DKT mechanism takes place. In this regime, the two bubbles deviate almost symmetrically from their initial paths and this deviation happens while they get very close to each other, the remaining gap being of the order of the bubble radius or even less. After the tumbling stage is completed, the two bubbles rise virtually side by side in a vertical plane, their horizontal centre-to-centre separation being approximately $3$ bubble radii. In more inertial regimes, say $12\lesssim Ga\leq30$, and low-to-moderate Bond numbers, the bubble pair evolves in such a way that the TB escapes laterally from the wake without significantly altering the path of the LB. In this Asymmetric Side Escape scenario, the lateral drift of the TB takes place while the two bubbles are still widely separated. In the subsequent stage, the tandem maintains a significant and almost constant inclination in the range $30^\circ\lesssim\theta\lesssim40^\circ$, except for near-coalescence conditions under which the side-by-side configuration may eventually be reached. Whatever $Ga$, coalescence is avoided only up to a critical Bond number, the value of which increases with $Ga$, from $Bo_c\approx0.1$ for $Ga=10$ to $Bo_c\approx0.5$ for $Ga=30$. Actually, bubble deformation influences the evolution of the system in a number of ways. In particular, the fact that the vorticity generated on a bubble is directly proportional to the curvature of its surface implies that the strength of the wake-induced interaction acting on the TB increases sharply with the Bond number. This interaction being attractive, deformation effects are found  to promote the suction of the TB along the centreline of the LB wake. This of course favours coalescence, but also tends to stabilize the in-line  configuration when $Bo<Bo_c(Ga)$ by delaying the growth of non-axisymmetric disturbances. \\
\indent We performed a detailed examination of the influence of initial conditions, especially of a possible misalignment of the two bubbles. Even a minimal initial deviation ($\theta_0\leq2^\circ$) was found to have a dramatic influence on the evolution of the bubble pair. In particular, the DKT regime previously observed for $Ga=\mathcal{O}(10)$ and $Bo\lesssim0.1$ no longer takes place. Instead the system follows an ASE evolution. A nonzero initial inclination also promotes the ASE configuration toward larger bubble deformations, making the critical Bond number increase up to $0.8$. The reason is that for $\theta_0\neq0^\circ$ the TB faces an asymmetric flow from the very beginning of its rise, which makes it able to drift laterally over a longer time than in the canonical $\theta_0=0^\circ$ case. This scenario is efficient to avoid coalescence as far as the sideways force is dominated by the classical shear-induced lift mechanism. However, mechanisms leading to a weakening or even a reversal of the overall transverse force exist for non-spherical bubbles. In particular, bubbles rising in a shear flow exhibit a non-axisymmetric shape, a feature known to produce a deformation-induced component of the transverse force with opposite sign compared to that of the shear-induced inertial lift. As the Bond number approaches its critical value, we observed that the asymmetry of the carrying flow in which the TB is immersed for $\theta_0\neq0^\circ$ yields a pronounced egg-like shape of the latter. This goes hand in hand with a reduction of the transverse force which eventually changes sign for $Bo>Bo_c$, forcing the two bubbles to coalesce. \vspace{1mm} \\
\indent The present study clarifies the respective roles of inertial, viscous and capillary effects, as well as that of initial conditions, in the three-dimensional dynamics of the considered system. 
From the standpoint of the microstructure of bubbly suspensions, the main outcome is presumably the final geometry of the arrangement that emerges from the evolution of bubble pairs initially released in line, possibly with some small angular deviation. As we saw, only pairs made of nearly-spherical bubbles with $\mathcal{O}(10)$-Galilei numbers end up in a side-by-side configuration and thereby tend to favour the formation of horizontal clusters. However, in most non-coalescing situations, the interaction process follows the ASE scenario, yielding final inclinations in the range $15^\circ-40^\circ$ and horizontal separations ranging from $2$ to $5$ radii, depending on $Ga, Bo$ and $\theta_0$. Given the unavoidable variations of initial bubble positions in real bubbly flows, this significant range of near-equilibrium inclinations and separations guarantees a much more homogeneous spatial bubble distribution on the long term than what can be expected from simplified models assuming potential flow and/or spherical bubble shapes. \\
\indent \color{black} This study did not examine the influence of slight differences in the size of the two bubbles. Nevertheless the main consequences of such a difference may be inferred from present findings. Suppose that the LB is slightly larger than the TB. In the $(Ga,\,Bo)$-range considered here, the corresponding increase in the buoyancy force makes the rise speed of the LB increase, lowering the positive velocity difference $V_{TB}-V_{LB}$ during the axisymmetric stages of the interaction process. Therefore, compared to the reference case, the two bubbles maintain a larger separation at a given time, which favours the occurrence of an ASE-type scenario. Consequently, this size difference tends to broaden the subdomain corresponding to the ASE regime in the phase diagram of figure \ref{f4.2.1}. Conversely, if the TB is slightly larger than the LB, the two bubbles are more prone to getting close to each other before the axial symmetry of the flow breaks down. Hence, if $Ga$ and $Bo$ are such that the system stands close to the DKT-ASE transition, this configuration favours the DKT scenario. Similarly, it lowers the critical Bond number $Bo_c(Ga)$ if the system is close to the coalescence threshold. The above predictions may be corroborated with the experimental observations of \cite{Kusuno2019} who considered two bubble pairs close to the DKT-ASE transition, with a LB corresponding to $Ga=13.5,\,Bo=0.27$ in both cases. With a TB $3.5\%$ larger than the LB, they found the system to follow a DKT transition (their figures 4$(c)$ and 5$(c)$). Conversely, they observed a clear ASE scenario when the TB was $3.5\%$ smaller than the LB (figures 4$(d)$ and 5$(d)$).\\
\indent \color{black} Another aspect that the present study leaves untouched is the fundamental question of the mathematical nature of the bifurcation that takes place when the axisymmetric configuration becomes unstable. Similarly, although the DKT and ASE regimes exhibit markedly different physical characteristics, whether or not they correspond to truly distinct unstable modes of the system remains unknown at this stage. Tackling these issues requires the development of an appropriate global linear stability approach. Numerical tools allowing the threshold and nature of bifurcations involved in the wake of fixed \citep{Tchoufag2013,Cano2016} or freely-moving \citep{Tchoufag2014} clean isolated bubbles with a prescribed shape have become available during the past decade. More recently, the same approach was extended to fully deformable isolated bubbles \citep{Bonnefis2019}. Further extending this approach to systems involving bubble pairs is certainly feasible and seems the natural next step capable of bringing new insight into these fundamental issues. 

\section*{Acknowledgements.}
 \noindent The authors acknowledge the partial support of this research by NSFC ( Natural Science Foundation of China) under grants $\#51636009,\, \#11872296$, and $\#U173227$. J. Z. acknowledges the Young Elite Scientists Sponsorship Program under grant YESS No. 2018QNRC001.
 \vspace{-5mm}
 \section*{Declaration of interests.}  
 \noindent The authors report no conflict of interest.
 \color{black}
 \appendix
 \section{Numerical tests}
 \label{anum}
 \begin{figure}
  \vspace{4mm}
  \centerline{\includegraphics[width=0.98\textwidth]{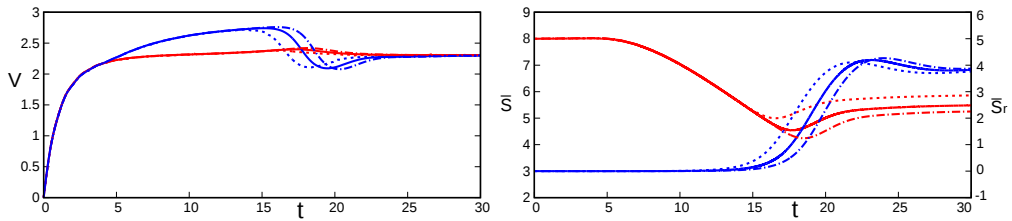}}
   \vspace{3mm}
  \caption{\color{black}Influence of the tolerance imposed to the Poisson solver on the evolution of a bubble pair with $Ga=30,\, Bo=0.3$. $(a)$: vertical velocity component of the LB (red line) and TB (blue line); $(b)$ vertical (red line, left axis) and horizontal (blue line, right axis) components of the separation. Dotted, solid and dash-dotted lines refer to simulations performed with a tolerance of $1\times10^{-2},\,1\times10^{-4}$ and $1\times10^{-6}$, respectively.   }
\label{toler}
\end{figure}
\begin{figure}
  \vspace{4mm}
  \centerline{\includegraphics[width=0.98\textwidth]{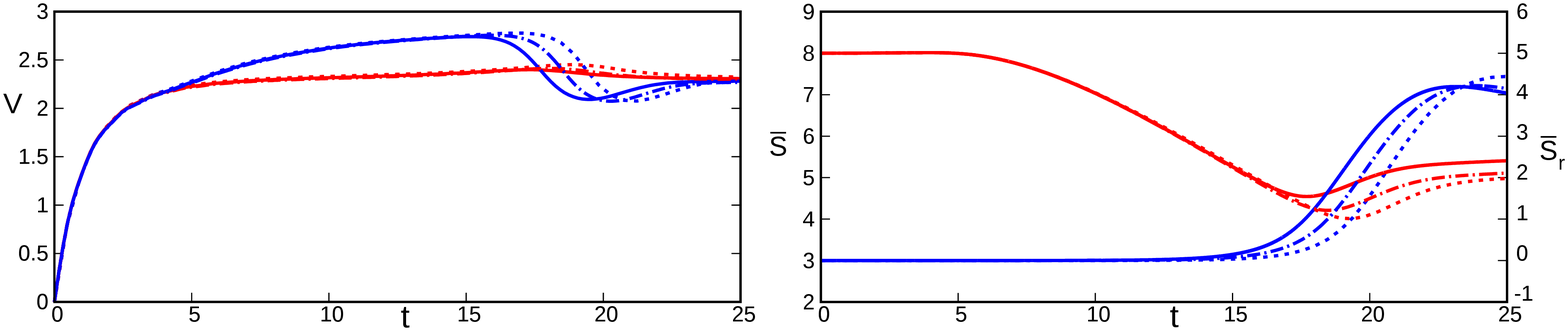}}
   \vspace{3mm}
  \caption{\color{black}Influence of grid resolution on the evolution of a bubble pair with $Ga=30,\, Bo=0.3$. $(a)$: vertical velocity component of the LB (red line) and TB (blue line); $(b)$ vertical (red line, left axis) and horizontal (blue line, right axis) components of the separation. Dotted, solid and dash-dotted lines refer to simulations performed with $\Delta_{min}=R/34,\,R/68$ and $R/136$, respectively.   }
\label{level}
\end{figure}
To assess the accuracy of the three-dimensional computations, two series of tests were performed, both with the parameters $(Ga=30,\,Bo=0.3)$ corresponding to the ASE configuration discussed in \S\,\ref{sec4.4}. To correctly interpret these tests, it is crucial to keep in mind that the initial configuration being axisymmetric, the transition to a non-axisymmetric state is governed by the asymmetry of numerical disturbances. \\
\indent The Poisson solver is the only part of the time-advancement algorithm used in \textit{Basilisk} that does not preserve spatial symmetry up to machine accuracy \citep{Popinet2003}. For this reason, we first examined how much the user-specified tolerance $T_\epsilon$ applied to this solver influences the evolution of the bubble pair. This tolerance is defined as the maximum relative change during one time step of the fluid volume enclosed in a cell, \textit{i.e.} the maximum of $|\nabla\cdot\textbf{\textit{u}}|\Delta t$ over the computational domain. The standard tolerance used throughout the computations is $T_\epsilon=1\times10^{-4}$. In this test, we ran two additional simulations, with $T_\epsilon=1\times10^{-2}$ and $1\times10^{-6}$, respectively. Figure \ref{toler} reveals that the smaller $T_\epsilon$ is, the longer it takes for the axial symmetry of the system to be broken (see the evolution of $\overline{S}_r$ in figure \ref{toler}$(b)$). This was to be expected since reducing the tolerance reduces the asymmetry of the solution, delaying the transition to three-dimensionality. This of course has some impact on the final vertical separation of the two bubbles, hence on the inclination of their line of centres. Conversely, the bubble velocities before and after the lateral escape of the TB are left unchanged by the change in $T_\epsilon$, together with the final horizontal separation.\\
\indent Then we carried out a grid convergence study, setting $T_\epsilon=1\times10^{-4}$ in all cases. Starting from the reference grid with $\Delta_{min}=R/68$ used throughout the study, we ran the same case on a grid twice coarser ($\Delta_{min}=R/34$) and another grid twice finer ($\Delta_{min}=R/136$). 
\color{black} Note that, since the time step decreases with $\Delta_{min}$, the computation with $\Delta_{min}=R/136$ was extremely expensive. This is why it was stopped at $t=25$, beyond which the dynamics of the flow is not expected to reveal any supplementary influence of the grid resolution. 
 The results of these tests are presented in figure \ref{level}. Again, these results show that the only significant change in the velocities evolution is the time by which the TB starts to escape laterally, and consequently the final vertical separation of the two bubbles (although only evolutions for $t\leq25$ are displayed in the figure, the horizontal separation was found to relax to the same value with the two grids on which the runs were carried out over larger times). It is noticeable that the onset of three-dimensionality is reached earlier on the reference grid than on both the finer and coarser ones. To understand this surprising feature, it must be kept in mind that the time step limitation arises from capillary effects, implying $\Delta t\propto \Delta_{min}^{3/2}$. Therefore, the maximum error on $|\nabla\cdot\textbf{\textit{u}}|$ resulting from the Poisson solver is proportional to $\Delta_{min}^{-3/2}$. Assuming that the cell size at the location where this maximum is reached is $\Delta$, the corresponding error on $\textbf{\textit{u}}$ is proportional to $\Delta_{min}^{-3/2}\Delta$. Since the local cell volume is $\Delta^3$, the contribution of the error to the local fluid momentum is proportional to $\Delta_{min}^{-3/2}\Delta^4$. Indeed, what determines the influence of the error on the onset of the lateral motion is the asymmetric component of the momentum over a cell rather than that of the local velocity $\textbf{\textit{u}}$ itself. The position of the maximum error within the computational domain is unknown \textit{a priori}. It may vary from one grid to another, and this variation is responsible for the non-monotonic behaviour observed in figure \ref{level}. For instance, if the maximum is reached in the most refined subregion (\textit{i.e.} very close to one interface) on the coarsest grid ($\Delta=\Delta_{min}=R/34$) while it is reached in the near wake on the other two grids ($\Delta=4\Delta_{min}=R/17$ and $R/34$, respectively), the asymmetric contribution to the momentum (normalized with $(R/34)^{5/2}$) is of $\mathcal{O}(1)$ on the coarsest grid, while it is of $\mathcal{O}(2^{11/2})$ and $\mathcal{O}(2^{3})$ on the intermediate and finest grids, respectively. In such a case, the onset of the TB lateral  escape is expected to happen first on the intermediate grid, then on the finest one, and finally on the coarsest one, just as observed in the figure. \\
\indent In summary, the above tests revealed that, as expected, the minimum cell size and tolerance on the Poisson solver influence the time by which the axial symmetry of the initial configuration breaks down. The smaller the tolerance $T_\epsilon$, the longer it takes for the bifurcation to take place on a given grid. The influence of the grid resolution is more complex, owing to the variability of the position at which the maximum asymmetry on the velocity field takes place. Because of this, the time by which the solution develops a significant three-dimensional component depends on both $\Delta_{min}$ and the local cell size $\Delta$ at the location of this maximum. Obviously, a longer transition time being equivalent to as shorter separation between the two bubbles, the sensitivity of the system to these numerical parameters implies that the value of the critical Bond number $Bo_c(Ga)$ has a nonzero numerical `error bar'. For instance, although the bubble pair with $(Ga=30, Bo=0.45)$ is found to escape coalescence with $\Delta_{min}=R/64$ and $T_\epsilon=1\times10^{-4}$, it is very likely that it coalesces with a smaller $T_\epsilon$. The numerical uncertainty on $Bo_c(Ga)$ may be estimated by considering the time lag for the onset of the TB lateral escape resulting from the sensitivity of the system to $T_\epsilon$ and $\Delta_{min}$. Given the difference in the rise speed of the two bubbles by the time this escape starts, figures \ref{toler} and \ref{level} indicate that the vertical separation between the two bubbles is approximately reduced by $\Delta\overline{S}=0.3$ when $T_\epsilon$ is reduced by two orders of magnitude or when the grid resolution is increased from $\Delta_{min}=R/68$ to $\Delta_{min}=R/136$. Then, figure \ref{f5.1.1}$(b)$ indicates that the minimum vertical separation reached during the lateral escape varies from $4.55$ for $Bo=0.3$ to $2.0$ for $Bo=0.45$, from which a rule of three suggests that a reduction of the vertical separation by $\Delta\overline{S}=0.3$ implies a reduction of the critical Bond number by $1.8\%$.
\color{black}
\color{black}
\section{Bubble coalescence in pure liquids}
\label{coal}
Coalescence of drops and bubbles in a suspending liquid has received considerable attention in the literature owing to its many applications. From the fluid mechanics viewpoint, most of the studies carried out during the second half of the past century attempted to determine the characteristics of the drainage of the film that forms in between two drops or bubbles when they approach each other (see \cite{Chesters1991} and \cite{Chan2011} for reviews). Assuming an axisymmetric geometry, asymptotic studies based on the lubrication approximation and numerical studies based on the boundary integral method (\textit{e.g.} \cite{Chi1989}) revealed the critical role of the drop/bubble-to-external fluid viscosity ratio in the drainage dynamics. Depending on how this ratio, $\lambda$ say, compares with the film aspect ratio $\epsilon$ (the typical radius-to-thickness ratio of the near-contact region), four possible situations arise, corresponding to nearly immobile ($\lambda\gg\epsilon^{1/2}$), partially mobile ($\lambda\sim\epsilon^{1/2}$), mobile ($\epsilon^{-1/2}\ll\lambda\ll\epsilon^{1/2}$) and fully mobile ($\lambda\ll\epsilon^{-1/2}$) interfaces, respectively \citep{Davis1989,Chesters1991}. In the first three cases, 
provided the drainage takes place under the action of a constant external force and the drops are nearly spherical except in the near-contact region, the minimum film thickness obeys a power law evolution \citep{Hartland1968,Jones1978, Yiantsios1989, Nemer2013}. The corresponding exponent depends on the flow and boundary conditions, but stands in between $0$ and $-1$ in all cases. A crucial consequence of this algebraic thinning rate is that the drainage requires an infinite time to be completed. Hence, non-hydrodynamic effects, in the first place the long-range London-van der Waals force, are required for coalescence to be achieved in a finite amount of time. \\
\indent Things differ drastically with fully mobile interfaces, which is the relevant situation for bubbles in pure liquids (contamination by surfactants yields immobile or at least partially mobile interfaces; \textit{e.g.} \cite{Vakarelski2019,Vakarelski2020}). In this case, the flow in the gap is merely a plug flow \citep{Davis1989}, as interfaces offer no resistance to the squeezing of the film. Under such conditions, standard lubrication approximations do not hold unless the gap has become extremely thin \citep{Davis1989,Yiantsios1989,Nemer2013}. The film thickness decreases exponentially over time during most of the drainage process, the decay rate depending on whether the Reynolds number $Re_a=\rho V_aR/\mu$ based on the relative approach velocity $V_a$ of the bubbles is large or small \citep{Chesters1991}. This exponential thinning law was confirmed experimentally \citep{Debregeas1998} and numerically \citep{Pigeonneau2011} by considering buoyancy-driven bubbles reaching a free surface. This law holds as far as the film thickness can be considered uniform. However, similar to the case of immobile or partially mobile interfaces, the minimum film thickness initially located on the line of centres shifts gradually to the film periphery, giving rise to a `dimple' corresponding to the transition region between the film and the outer flow. To take the influence of this dimple into account, \cite{Chesters1982} solved numerically the set of thinning equations with appropriate boundary conditions for initially spherical bubbles, assuming $Re_a\gg1$, \textit{i.e.} considering that inertial and capillary effects are in balance. They observed that, when the minimum film thickness has reduced sufficiently, the thinning velocity at the dimple position levels off at a value close to $0.1V_a$. Under constant-force conditions, the crucial consequence of this thinning evolution is that drainage is completed in a finite time without the need for a non-hydrodynamic force to intervene.  Defining the approach Weber number $We_a=\rho V_a^2R/\gamma$, the dimensionless inertial drainage time is then 
\begin{equation}
\overline{T}_{di}=k_iWe_a/\overline{V}_a\,,
\label{CH}
\end{equation}
 with $k_i\approx1.0$ and $\overline{V}_a=V_a/(gR)^{1/2}$ in the buoyancy-driven case of interest here. The value of $k_i$ was later slightly re-evaluated to $k_i\approx1.08$ by \cite{Duineveld1994,Duineveld1998}. Within a liquid film bounded by two clean gas-liquid interfaces, the London-van der Waals force is attractive, thus shortening the coalescence time. However this force is known to be significant only when the distance between the two interfaces is less than $100\,$nm, which made Chesters \& Hofman conclude that it barely shortens the coalescence process.\\
\indent Obviously, coalescence occurs only if the relative approach velocity remains positive throughout the drainage. For this to be the case, only part of the kinetic energy resulting from the relative motion of the two bubbles (moving with velocities $\pm V_a/2$) must be converted into surface energy through the deformation of the bubble-fluid interface in the near-contact region. Still in the limit $Re_a\gg1$, this criterion yields a critical Weber number $We_{ac}$ beyond which the two bubbles bounce once or several times, until eventually coalescing when the approach Weber number has decreased sufficiently. Under potential flow assumptions, the kinetic energy associated with the bubble relative motion is proportional to the virtual mass (or added-mass) coefficient $C_{Ma}$ in the corresponding direction, making $We_{ac}$ vary linearly with $C_{Ma}$. For nearly-spherical bubbles, \cite{Chesters1982} established that the relative increase of the surface energy is at leading order $(k_iWe_a/4)^2$, which yields $We_{ac}=\frac{2}{3}k_i^{-2}C_{Ma}$. For two spheres in contact after having moved toward each other , $C_{Ma}\approx0.80$ \citep{Voinov1969,Miloh1977}, so that $We_{ac}\approx0.45$ \citep{Duineveld1994,Duineveld1998}. \\
\indent In addition, Chesters \& Hofman pointed out that previous results also apply to a single bubble reaching a free surface, up to a simple geometric transformation. The outcome is that in this configuration (\ref{CH}) transforms into $\overline{T}_{di}=4k_iWe_a/\overline{V}_a$, while the above criterion for the onset of bouncing becomes $We_{ac}=\frac{1}{3}k_i^{-2}C_{Ma}$. Assuming the bubble to be spherical and the Froude number of the free surface, $\overline{V}_a^2$, to be large, the relevant virtual mass coefficient is $C_{Ma}\approx0.42$ \citep{Miloh1977}, so that $We_{ac}\approx0.12$ \citep{Duineveld1994}. By tracking sub-millimeter size bubbles rising up to a free surface in ultrapure water, the same author determined the experimental threshold as $We_{ac}\approx0.105$ (corresponding to a rise Reynolds number $Re\approx50$). This agreement with the theoretical prediction is supported by more recent experiments \citep{Vakarelski2020} and provides an important support to the approach of \cite{Chesters1982}, although some of its aspects have been questioned \citep{Chan2011}. \\
\indent The above conclusions hold for nearly-spherical bubbles provided effects of the liquid viscosity are negligible. However, despite the uniform velocity profile in the film, viscous effects arise through normal stresses. For a given film thickness, these effects tend to increase the film radius, \textit{i.e.} the area of the near-contact region. This results in a significant decrease of the film thinning rate as soon as $Re_a\lesssim10$ \citep{Chesters1982}. In this $Re_a$-range, this thinning rate decreases continually as the drainage proceeds, making the London-van der Waals force inescapable for coalescence to occur.   
For low enough $Re_a$, the film evolution is governed by a viscous-capillary balance, hence by the capillary number $Ca_a=We_a/Re_a=\mu V_a/\gamma$. To the best of our knowledge, no theoretical model is available to predict the drainage time in this viscosity-dominated regime for fully mobile interfaces. However, recent experimental data may be used to obtain an empirical scaling law from which realistic estimates may be inferred. The case of air bubbles coalescing at a clean free surface after rising in a liquid $20$ times more viscous than water was considered in detail by \cite{Vakarelski2018}. Small bubbles ($R\lesssim0.1$\,mm) were observed to coalesce almost immediately on their arrival at the surface, while larger bubbles remained `glued' there during some time without bouncing, until coalescence eventually occurred. A single tiny bounce was detected for bubbles larger than $R\approx0.45$\,mm, corresponding to $We_a\gtrsim0.135$. 
Overall, they found the residence (\textit{i.e.} drainage) time $T_{dv}$ of the bubble at the surface, to be proportional to $R^2$. Their results may be used as a basis to derive a generic empirical expression for the viscous drainage time, $T_{dv}$. Considering that $\overline{T}_{dv}=(g/R)^{1/2}T_{dv}$ is primarily driven by the capillary number in this $\mathcal{O}(1)$-Reynolds number range and keeping in mind that the rise speed also grows like $R^2$ in this regime, it turns out that the simplest admissible scaling law for $\overline{T}_{dv}$ is
\begin{equation}
\overline{T}_{dv}=k_vCa_a^{3/2}/\overline{V}_a\,,
\label{Visc}
\end{equation}
with $k_v=1.8\times10^3$ according to the above experimental data.
It is then a simple matter to compare the inertial and viscous estimates for the drainage time in a given fluid and for a given bubble size. For instance, in ultrapure water ($Mo=2.6\times10^{-11}$), the bounce/no bounce threshold for a bubble reaching a free surface is known to correspond to $0.335$\,mm-radius bubbles \citep{Duineveld1994}. Under such conditions, (\ref{CH}) (with the appropriate transformation) and (\ref{Visc}) yield $\overline{T}_{di}\approx0.32$ and $\overline{T}_{dv}\approx0.064$, respectively. Hence the drainage is controlled by inertial effects and coalescence takes place very soon after the bubble collides with the free surface. Similarly, in the viscous liquid used by \cite{Vakarelski2018} ($Mo=6.6\times10^{-5}$), the same two predictions for a bubble with $R=0.45\,$mm (\textit{i.e.} just below the bounce/no bounce threshold) yield $\overline{T}_{di}\approx0.68$ and $\overline{T}_{dv}\approx23.8$, respectively. In this case, the drainage time estimated through the viscous law (\ref{Visc}) is $35$ times larger than that predicted by the inviscid approach, implying that the latter is irrelevant. This corresponds to a situation in which the bubble stays `glued' to the free surface during a long time before coalescing. For a pair of bubbles rising in line under similar conditions, the two bubbles form a compound `dumbbell' bubble during the drainage process, as observed by \cite{Sanada2006} and \cite{watanabe2006line}. To estimate $T_{dv}$ in this case, (\ref{Visc}) must be modified according to the geometric transformation of Chesters \& Hofman, which yields $\overline{T}_{dv}=\frac{1}{2}k_vCa_a^{3/2}/\overline{V}_a$.\\
\indent Most results reviewed above were obtained with nearly-spherical bubbles ($Bo\ll1$, in practice $Bo\lesssim0.2$). Influence of a significant distortion of the bubble shape on the coalescence process is complex because it involves antagonistic effects. On the one hand, the curvature of the near-pole region of an oblate bubble is smaller than that of a spherical bubble, making the area of the near-contact region larger. Thus the radial position of the dimple shifts outward, and a longer time is required to squeeze the film with a given approach velocity. \cite{Duineveld1994} solved the set of inviscid thinning equations for oblate bubbles with various aspect ratios. Compared to the reference case, his results show that the pre-factor $k_i$ involved in the prediction (\ref{CH}) for the drainage time is increased approximately by a factor of $2$ (resp. $3$) for $\chi=1.5$ (resp. $\chi=2$). 
 On the other hand, an oblate body moving along its short axis displaces more fluid than a sphere, which translates into a larger virtual mass coefficient, hence a larger kinetic energy available for the drainage. Moreover, as explained in \S\,\ref{mechanisms}, the bubble oblateness enhances wake effects, increasing the amount of fluid displaced by the bubble through the entrainment process in the wake. Therefore, for a given approach velocity, the kinetic energy of the fluid displaced by oblate bubbles may be significantly larger than the estimate based on the simple irrotational added-mass concept. Unfortunately, no drainage time prediction incorporating wake entrainment and/or viscous effects seems available for such bubbles.

\color{black}

\section{\color{black}Lift reversal mechanisms on distorted bubbles}
\label{vortrev}
As is well known, the shear-induced lift mechanism mentioned in \S\,\ref{mechanisms}, hereinafter called $L$-mechanism, stems from the bending of the vorticity of the carrying shear flow past the bubble, a process that results in the formation of a pair of counter-rotating streamwise vortices in its wake \citep{legendre1998lift}. The vorticity produced at the bubble surface plays no role in this mechanism which is inviscid by nature  \citep{Lighthill1956,auton1987lift}. However, for reasons discussed in \S\,\ref{mechanisms}, the strength of the surface vorticity resulting from finite-$Re$ effects increases sharply with the bubble oblateness. When an oblate bubble rises in a fluid at rest, the amount of vorticity produced at its surface becomes large enough for the axisymmetric wake to become unstable within a finite range of Reynolds number, $Re_{S-}(\chi)\leq Re\leq Re_{S+}(\chi)$, provided the bubble aspect ratio exceeds a critical value $\chi_{cS}\approx2.2$ \citep{magnaudet2007wake}. This mechanism is at the root of the path instability of millimeter-size bubbles rising in water \citep{mougin2001path}. Similar to the above $L$-mechanism, it gives rise to a wake dominated by a pair of counter-rotating streamwise vortices, the sign of which is selected by some initial disturbance. This wake instability mechanism, hereinafter called $S$-mechanism, subsists when a strongly oblate bubble rises in a weak shear flow. The only difference is that the initial disturbance is now provided by the outer shear, so that the sign of the streamwise vorticity in each trailing vortex is no longer random. Rather, a detailed analysis reveals that the contributions of the $L$- and $S$-mechanisms to the vortex tilting term involved in the streamwise vorticity balance have opposite signs \citep{adoua2009reversal}. For this reason, the sign of the overall sideways force depends on the relative strength of the two mechanisms. If the outer shear is strong enough, the $L$-mechanism dominates even for $\chi>\chi_{cS}$ and the lift force keeps the sign it would have for a spherical bubble. In contrast, if the shear is weak enough and $Re$ and $\chi$ fall in the range where the $S$-mechanism is active, the latter becomes dominant, yielding a reversed lift force. The lower Reynolds number beyond which lift reversal governed by the $S$-mechanism takes place in a weak shear flow is a sharply decreasing function of $\chi-\chi_{cS}$, with $Re_{S-}\approx155$ for $\chi=\chi_{cS}$ and $Re_{S-}\approx55$ for $\chi=2.5$ for instance. Conversely, the upper Reynolds number beyond which the lift force recovers the sign predicted by the $L$-mechanism dramatically increases with $\chi-\chi_{cS}$, from  $Re_{S+}=Re_{S-}\approx155$ for $\chi=\chi_{cS}$ to $Re_{S+}\approx680$ for $\chi=2.5$ for instance \citep{adoua2009reversal}. Because of this mechanism, one can suspect that bubbles experiencing a sufficient deformation may render the in-line configuration stable. Indeed, if the $S$-mechanism dominates, any deviation of the TB from the wake axis is expected to be counteracted by the reversed lift force.\\
\indent A distinct mechanism may also lead to the same effect on a non-axisymmetric bubble (more generally a drop). This mechanism, hereinafter called $A$-mechanism, is a consequence of the asymmetric deformation experienced by a drop immersed in a shear flow. In the zero-$Re$ limit, the drop is known to deform in such a way that its major and minor axes align with the eigen-directions of the associated strain rate \citep{Taylor1932}. In the case the drop has an additional translation with respect to the fluid, this deformation induces a nonzero sideways force, even at $Re=0$. When the Reynolds number is finite, this deformation-induced transverse force combines with the inertial shear-induced lift. However, theoretical predictions indicate that the two mechanisms result in sideways forces of opposite signs, except for drops whose viscosity is close to that of the suspending fluid (see equation (49) in \cite{magnaudet2003drag}). The deformation-induced (resp. inertia-induced) force being proportional to the Weber (resp. Reynolds) number, the direction of the total lift force is governed by the capillary number $Ca=We/Re$. This force changes sign for a critical value $Ca=Ca_c$ which, for a bubble with negligible internal viscosity, depends only on the relative shear rate $\beta R/u_T$, $\beta$ denoting the ambient shear rate. While the total lift force keeps the sign corresponding to the inertial shear-induced mechanism if $Ca<Ca_c$, it acts in the opposite direction for larger $Ca$. In the case of a bubble, this mechanism results from the fact that the non-penetration condition and the normal stress balance have to be jointly satisfied at the gas-liquid interface. This requirement obviously subsists for Reynolds numbers larger than unity and so does the above mechanism, although inertial effects tending to make the bubble oblate combine with those of the outer shear to produce more complex asymmetric shapes compared to the low-$Re$ configuration. Lift reversal due to the $A$-mechanism has been observed in computations \citep{ervin1997rise, Sankaranarayanan2002} and experiments performed with isolated bubbles rising in viscous liquids sheared in a Couette device \citep{tomiyama2002transverse,aoyama2017lift}.

\bibliographystyle{jfm}
\bibliography{bubble-inline}


\end{document}